\RequirePackage{fix-cm}
\documentclass[natbib,smallextended]{svjour3}       
\smartqed  
\usepackage{graphicx,amsmath,amssymb}
\usepackage{color}
\usepackage{aas_macros}
\usepackage{hyperref}
\newcommand{\simgt}{\lower.5ex\hbox{$\; \buildrel > \over \sim \;$}}
\newcommand{\simlt}{\lower.5ex\hbox{$\; \buildrel < \over \sim \;$}}

\def\h70Msol{\mathrel{h_{70}^{-1}M_\odot}}

\begin{document}

\title{Constraining Gas Motions in the Intra-Cluster Medium 
}


\author{ Aurora Simionescu \and
		John ZuHone \and 
        Irina Zhuravleva \and  
        Eugene Churazov \and
        Massimo Gaspari \and
  	Daisuke Nagai \and
        Norbert Werner \and	 
        Elke Roediger \and
        Rebecca Canning  \and
        Dominique Eckert \and
        Liyi Gu \and
        Frits Paerels            
}


\institute{Aurora Simionescu \at
SRON, Netherlands Institute for Space Research, Sorbonnelaan 2, 3584 CA Utrecht, The Netherlands; E-mail: a.simionescu@sron.nl \\
Institute of Space and Astronautical Science (ISAS), JAXA, 3-1-1 Yoshinodai, Chuo-ku, Sagamihara, Kanagawa, 252-5210, Japan
	\and
John ZuHone \at
Harvard-Smithsonian Center for Astrophysics, 60 Garden St., Cambridge, MA 02138, USA\\
	\and
Irina Zhuravleva \at
Department of Astronomy \& Astrophysics, University of Chicago, 5640 S Ellis Ave, Chicago, IL 60637, USA\\
Kavli Institute for Particle Astrophysics and Cosmology, Stanford University, 452 Lomita Mall, Stanford, CA 94305-4085, USA \\
Department of Physics, Stanford University, 382 Via Pueblo Mall, Stanford, CA 94305-4085, USA \\
	\and
Eugene Churazov \at
Max Planck Institute for Astrophysics, Karl-Schwarzschild-Strasse 1, D-85741 Garching, Germany \\
Space Research Institute (IKI), Profsoyuznaya 84/32, Moscow 117997, Russia \\
	\and    
Massimo Gaspari \at
Einstein and Spitzer Fellow,
Department of Astrophysical Sciences, Princeton University, 4 Ivy Lane, Princeton, NJ 08544-1001, USA
	\and
Daisuke Nagai \at
Department of Physics, Yale University, PO Box 208101, New Haven, CT, USA \\
Yale Center for Astronomy and Astrophysics, PO Box 208101, New Haven, CT, USA
	\and
Norbert Werner \at
MTA-E\"otv\"os Lor\'and University Lend\"ulet Hot Universe Research Group, H-1117 P\'azm\'any P\'eter s\'eta\'ny 1/A, Budapest, Hungary \\
Department of Theoretical Physics and Astrophysics, Faculty of Science, Masaryk University, Kotl\'arsk\'a 2, Brno, 61137, Czech Republic \\
School of Science, Hiroshima University, 1-3-1 Kagamiyama, Higashi-Hiroshima 739-8526, Japan \\
	\and
Elke Roediger \at
E.A. Milne Centre for Astrophysics, Department of Physics and Mathematics, University of Hull, Hull, HU6 7RX, UK
	\and
Rebecca Canning \at
Kavli Institute for Particle Astrophysics and Cosmology, Stanford University, 452 Lomita Mall, Stanford, CA 94305-4085, USA \\
Department of Physics, Stanford University, 382 Via Pueblo Mall, Stanford, CA 94305-4085, USA \\
	\and
Dominique Eckert \at
Max-Planck-Institut f\"ur extraterrestrische Physik, Giessenbachstrasse 1, 85748 Garching, Germany \\
Department of Astronomy, University of Geneva, Ch. d’Ecogia 16, CH-1290 Versoix, Switzerland
	\and 
Liyi Gu \at
RIKEN High Energy Astrophysics Laboratory, 2-1 Hirosawa, Wako, Saitama 351-0198, Japan \\
SRON Netherlands Institute for Space Research, Sorbonnelaan 2, 3584 CA Utrecht, the Netherlands \\
	\and
Frits Paerels \at
Columbia Astrophysics Laboratory and Department of Astronomy, Columbia University, New York, NY, USA
}

\date{Received: date / Accepted: date}

\maketitle

\begin{abstract}

The detailed velocity structure of the diffuse X-ray emitting intra-cluster medium (ICM) remains one of the last missing key ingredients in understanding the microphysical properties of these hot baryons and
constraining our models of the growth and evolution of structure on the largest scales in the Universe.  
Direct measurements of the gas velocities from the widths and shifts of X-ray emission lines were recently provided for the central region of the Perseus Cluster of galaxies by \textit{Hitomi}, and upcoming high-resolution X-ray microcalorimeters onboard \emph{XRISM} and \emph{Athena} are expected to extend these studies to many more systems. In the mean time, several other direct and indirect methods have been proposed for estimating the velocity structure in the ICM, ranging from resonant scattering to X-ray surface brightness fluctuation analysis, the kinematic Sunyaev-Zeldovich effect, or using optical line emitting nebulae in the brightest cluster galaxies as tracers of the motions of the ambient plasma.
Here, we review and compare the existing estimates of the velocities of the hot baryons, as well as the various overlapping physical processes that drive motions in the ICM, and discuss the implications of these measurements for constraining the viscosity and identifying the source of turbulence in clusters of galaxies.

\end{abstract}

\section{Introduction and motivation}

The evolution of structure in the Universe is a dynamical process. 
Both on the largest scales, where mergers between clusters of galaxies and accretion from the surrounding cosmic web drive the growth of the most massive haloes, and on smaller scales, where supermassive black holes (SMBHs) are intimately interconnected with the evolution of their host galaxies, information about the kinematics associated with these processes is an important component of astrophysical studies. 
Despite this fact, in particular when it comes to the hot, diffuse, X-ray emitting intergalactic medium (IGM), where most baryons in the local Universe reside, current observational results largely yield only a static picture, as if reducing a more complex movie to a single freeze-frame. On these very large scales, it would take tens of millions to billions of years between observations to actually notice a change of structure, so dynamical information can only be obtained through measurements of the velocities in the IGM. The best way to perform these measurements directly is to obtain X-ray spectra with a resolution better than at least a few eV; the \textit{Hitomi} satellite achieved a break-through in this field by providing spectra of this quality for the central region of the Perseus Cluster of galaxies. Upcoming X-ray observatories promise to expand our detailed knowledge of the intra-cluster medium (ICM), the densest, brightest parts of the IGM, to a much larger sample of systems. In parallel, other observational methods based on X-ray imaging, optical or sub-millimetre spectroscopy, and the thermal and kinematic Sunyaev-Zel'dovich (SZ) effect have also been proposed as alternative tests for the strength of motions in the ICM, leading up to these future direct constraints from X-ray spectroscopy.  

The goal of this review is to compile the existing quantitative constraints on the velocities of the intracluster plasma, and discuss their physical interpretation. We begin by summarising all the various processes that are expected to drive motions in the ICM from the perspective of theory and hydrodynamical simulations in Section \ref{sect_theor}, focusing on the expected magnitude, properties, and observational signatures specific to the velocity fields induced by cosmological large-scale structure formation (Section \ref{sect_lss}), idealised simulations of gas sloshing and ram-pressure stripped tails from subhalo infall (Sections \ref{sec:sect_cf} and \ref{sec:tails}), feedback from central active galactic nuclei (AGN; Section \ref{sec:AGNfeedback}), and other effects such as plasma instabilities and motions of the member galaxies (Section \ref{sect_instab}). We then introduce various methods of determining the level of gas motions, and their results, in Section \ref{sec:obs_probes}. Section \ref{sect_implic} discusses the interpretation of the inferred gas velocities in the context of identifying the source(s) of turbulence and constraining the microphysical properties of the ICM. Finally, in Section \ref{sect_future} we present the prospects of upcoming experiments that are expected to revolutionise this field of research. 

In general, the current manuscript is not intended as a theoretical overview of the physics of turbulence in a magnetised plasma beyond the MHD view, for which we refer the reader to, e.g. \citet{schekochihin06} for an introduction. There are many aspects of cosmological simulations of large-scale structure formation, the physics of gas sloshing, and AGN feedback that are not covered in this work; we have chosen to focus exclusively on how these processes affect the velocity structure of the ICM. Moreover, we only discuss the observational methods that provide independent constraints of the gas motions in individual systems, where the link to the dynamical state and disturbances in the gas density can be correlated in detail with the measured velocities to infer the physical processes that shape the kinematics of the ICM. The relevance of gas motions from the point of view of introducing deviations from the hydrostatic equilibrium (HSE) assumption and consequently influencing the measurement of cluster masses relevant for cosmological studies are addressed in the chapter by Pratt et al. in this volume.

\section{What are the physical processes that drive gas motions in the intergalactic medium?}\label{sect_theor}

\begin{figure}[t]
  \includegraphics[height=1.9in]{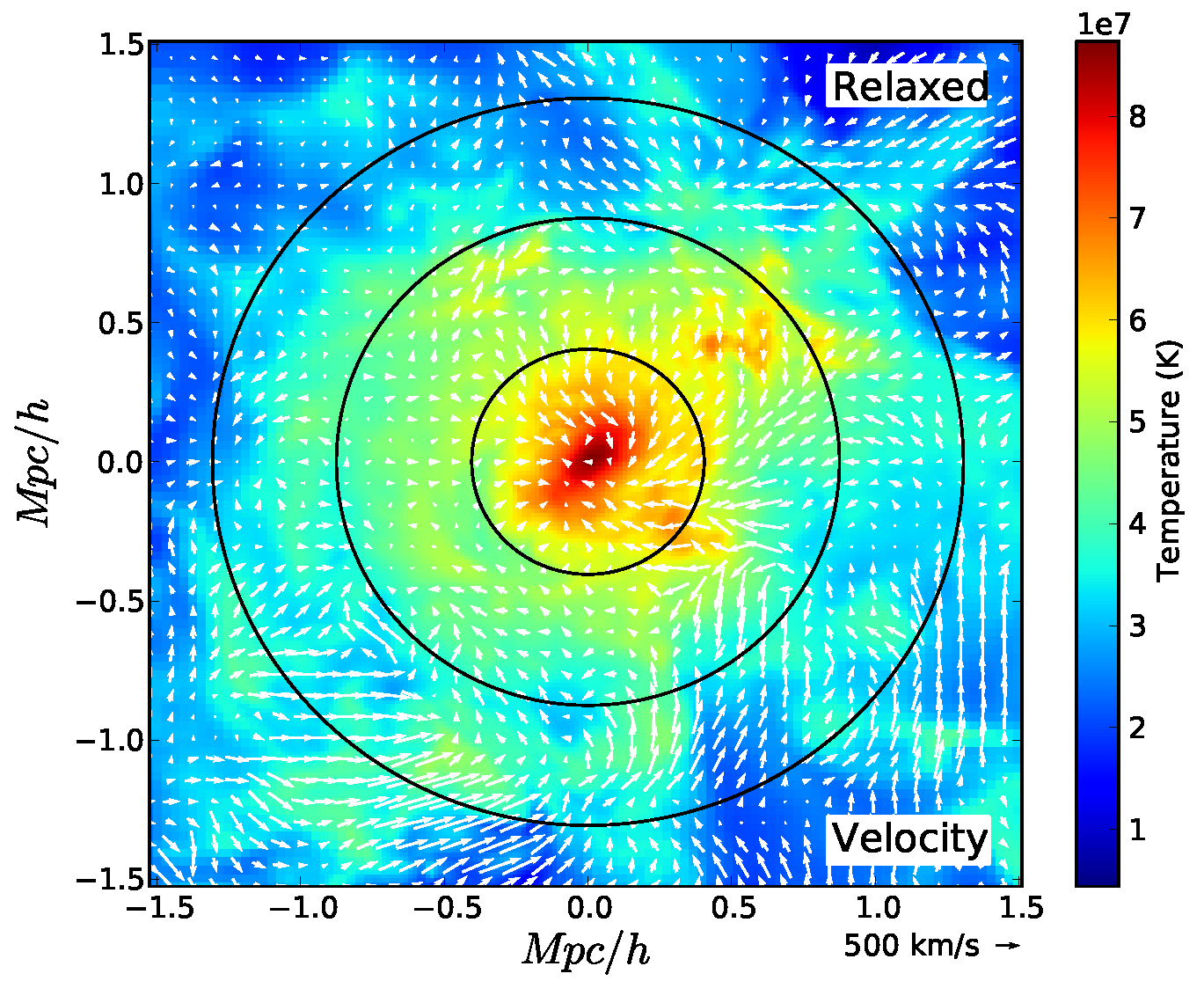}
  \includegraphics[height=1.9in]{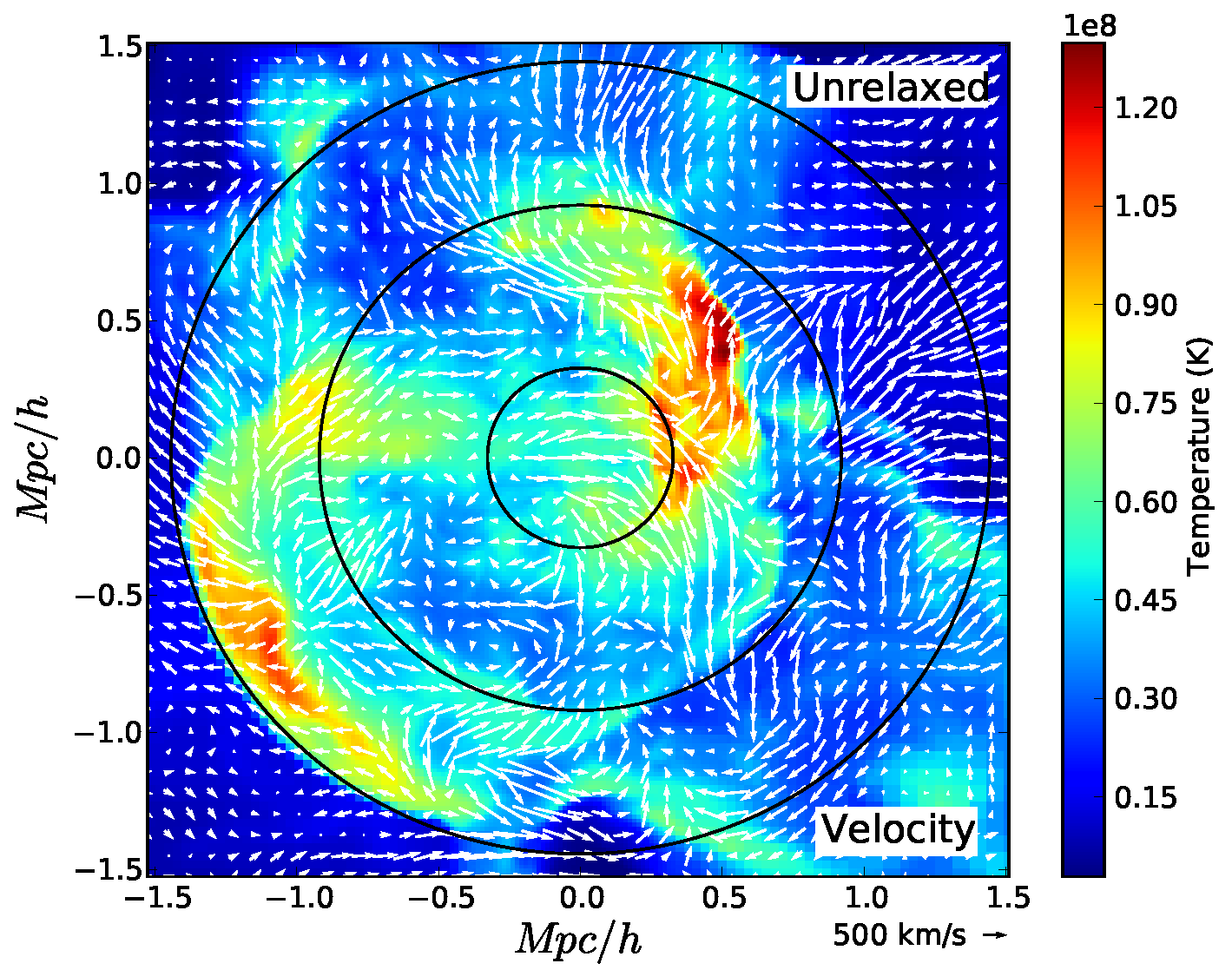}
  \caption{Projected mass-weighted temperature map of a relaxed (left) cluster and an unrelaxed (right) cluster with the velocity vector fields overlaid from the {\it Omega500} cosmological cluster simulation. The black circles denote $R_{2500}$, $R_{500}$ and $R_{200}$ of the clusters from inside to outside. Both the maps and vector fields are mass weighted along a $200$~kpc/h deep slice centered on their respective cluster centers. Reproduced from \citet{nelson14a}.}
\label{fig:turb-cosmo}
\end{figure}

\subsection{Large-scale structure formation} \label{sect_lss}

In the hierarchical structure formation model, clusters of galaxies form through a sequence of mergers and continuous mass accretion \citep{kravtsov12}.  These merging and accretion events generate a significant level of gas motions inside the cluster potential well, which eventually heat the gas through shocks or turbulent dissipation (Figure~\ref{fig:turb-cosmo}).  Hydrodynamical cosmological simulations of galaxy cluster formation using both grid-based \citep[e.g.,][]{norman99, ryu03, lau09, vazza09, vazza11, iapichino11, nagai13, miniati14} and particle-based \citep[e.g.,][]{dolag05b, battaglia12} methods have found that the intracluster gas motions generated in the structure formation process are ubiquitous and contribute significantly to the energy and pressure budget throughout the ICM \citep{lau09,nelson12,zhuravleva2013,nelson14b,shi15,shi18}. The ratio of non-thermal pressure due to these gas motions with respect to the thermal pressure increases, on average, from $\sim$10\% at $R_{500c}$ to $\sim$30\% at $R_{200c}$ \citep{lau09,vazza09,battaglia12}, while the cluster core physics (such as radiative cooling and feedback from stars and AGN) drives additional gas motions in cluster cores \citep[see Fig.~4 in][and \S\ref{sec:AGNfeedback} on the AGN feedback effect]{nagai13}.

It is thought that gas motions are the dominant form of non-thermal pressure in galaxy clusters, with current radio and gamma-ray observations limiting the contribution from other sources (cosmic-rays and the magnetic field) at only a few percent of the thermal energy content in the virialized regions of these systems \citep[e.g.][]{Ackermann10,Huber13,prokhorov2014,ackermann2014}\footnote{We note here that the gamma-ray and radio-based constraints might be weaker if the energy distribution of relativistic particles is softer than explicitly or implicitly assumed in these studies.}. Since X-ray and SZ observations typically measure only the thermal pressure component of the ICM, non-thermal pressure, if neglected, can introduce biases in the total energetics of the ICM as well as the hydrostatic mass estimation \citep[e.g.,][]{rasia06, nagai07,lau09,zhuravleva2013, biffi16, shi16}. For further detail, we refer the reader to the related chapter on ``The galaxy cluster mass scale and its impact on cosmological constraints from the cluster population'' (Pratt et al., submitted), in this topical collection.

\subsection{Cold fronts}\label{sec:sect_cf}

A particularly interesting manifestation of the ubiquitous motions associated with large-scale structure growth is represented by so-called ``cold fronts''. These sharp edges in X-ray surface brightness, where the brighter and denser side of the edge is colder than the other, were first discovered by \textit{Chandra} \citep{markevitch2000,vikhlinin2001} and are now known to be present in a large fraction of galaxy clusters \citep[see the reviews by][]{MV07,ZR16}.
Cold fronts occur in major mergers, such as in the ``Bullet Cluster'' \citep{markevitch2002} and Abell 3667 \citep{vikhlinin2001}, where the cold front is the contact discontinuity between the atmospheres of both merging subclusters. Cold fronts can also be produced by ``sloshing'' gas motions in the cool cores of clusters, triggered by perturbing subclusters, i.e., a minor merger \citep{markevitch2001,AM06}. In this case the contact discontinuity is not between two parcels of gas initially belonging to different systems (as in the case of major mergers), but is due to gas from the cool core of a cluster being moved by the spiral-shaped sloshing motion out to a larger radius where it encounters ICM with a different entropy, temperature, metallicity, and density \citep[e.g.][]{Tanaka2006,dupke2007b,Randall2009,Blanton2011}. Idealised simulations of binary minor cluster mergers, tailored to specific sloshing clusters in terms of the gravitational potential and ICM density and temperature profiles, can reproduce not only the locations of the observed sloshing cold fronts, but also the density and temperature contrasts across them (e.g., in Virgo and Abell 496, \citealt{Roediger2011,Roediger2012a496}). Thus, a careful interpretation of observed sloshing cold fronts, aided by tailored simulations, can be used to indirectly infer the current merger stage and the level of bulk motions. In addition, \citet{zinger2018} showed that sloshing can be triggered also by penetrating gas streams arising from cosmic accretion.
 Though usually seen near the centers of clusters, a number of deep X-ray observations have even revealed indications of cold fronts at radii of up to $\sim$1~Mpc \citep{simionescu2012,simionescu2017,paternomahler2013,rossetti2013,walker2014,walker2018Nat}. In all cases, the fronts are contact discontinuities generated by subsonic gas motions. Numerical simulations suggest these motions have Mach numbers of ${\cal M} \sim 0.3-0.5$, corresponding to velocities of several hundred km~s$^{-1}$. 

\begin{figure}
\begin{center}
\includegraphics[width=0.48\textwidth]{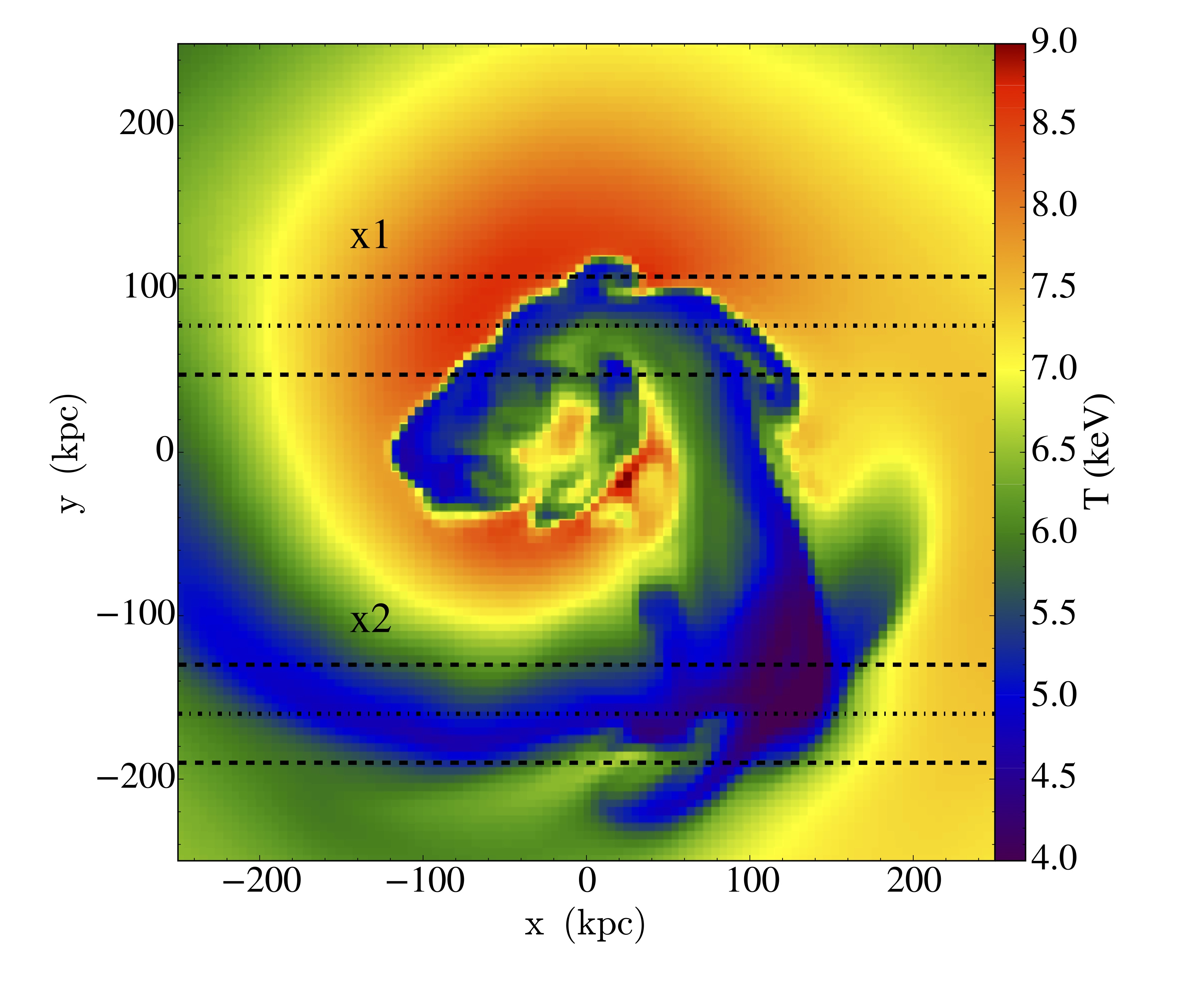}
\includegraphics[width=0.48\textwidth]{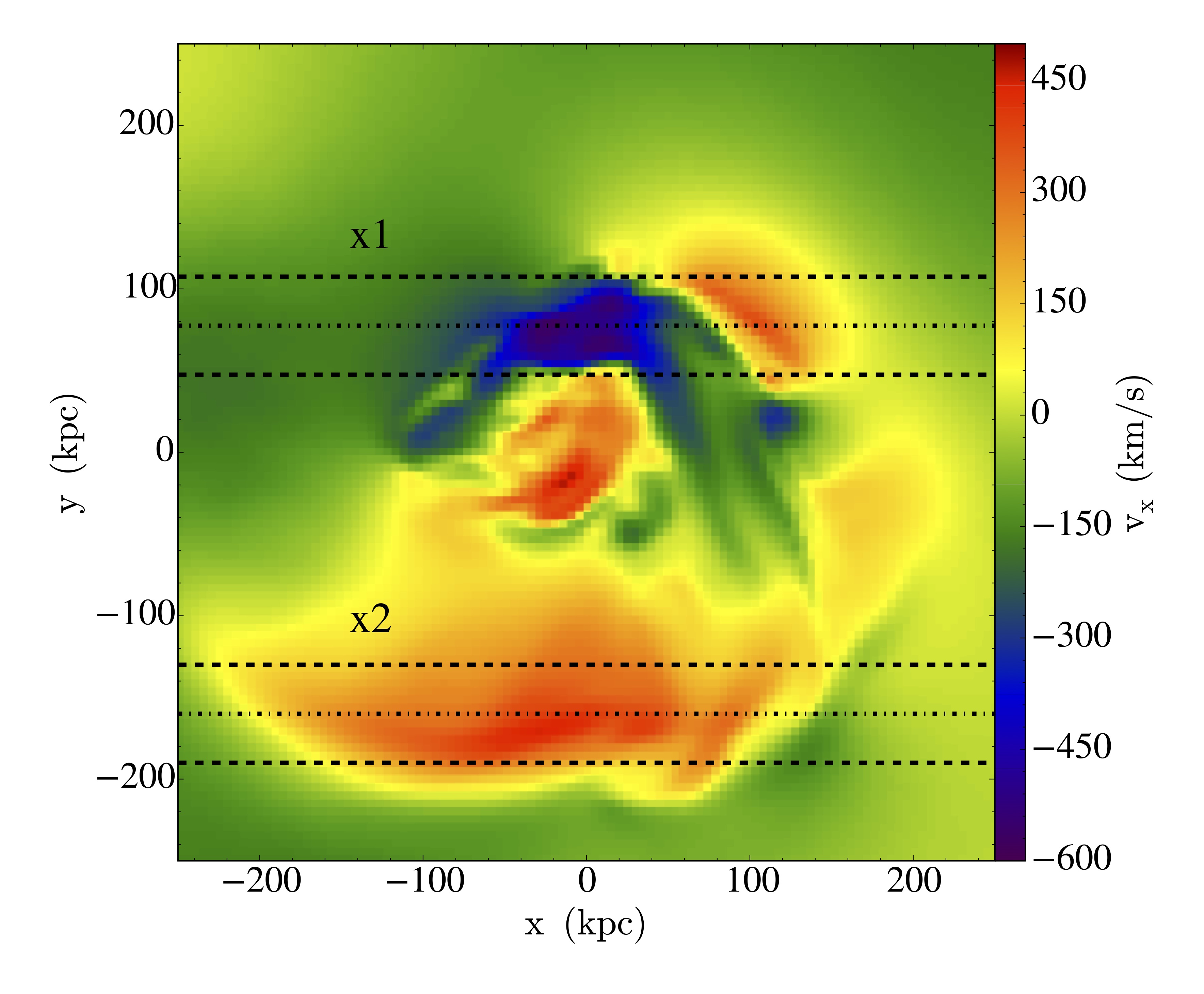}
\includegraphics[width=0.95\textwidth]{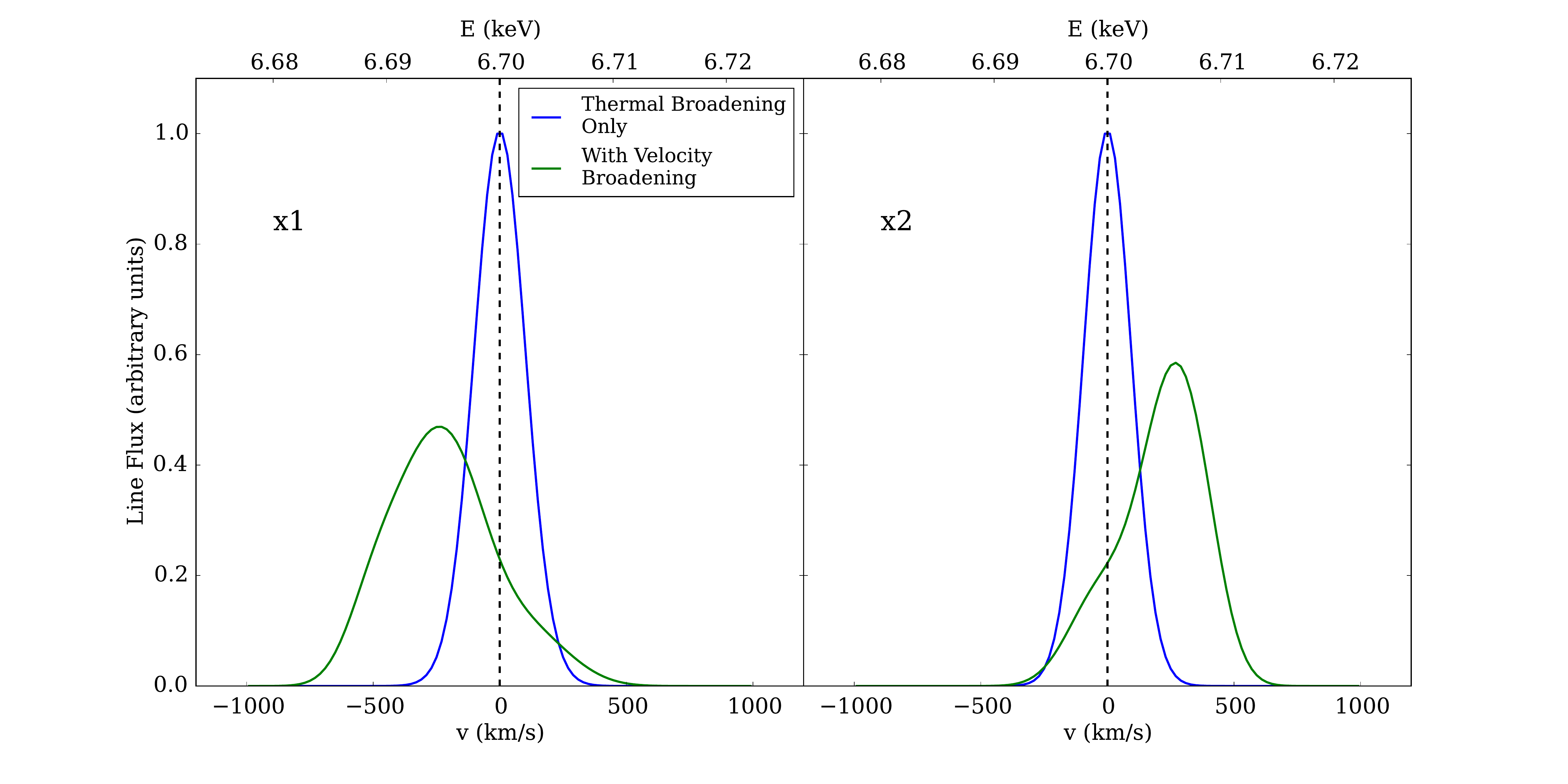}
\caption{Velocity fields of sloshing cold fronts from \citet{zuh2016}. Top panels: slices of temperature (left) and velocity (right) with regions of interest marked by dashed and dotted lines. Bottom panels: shifting (mean velocity) and broadening (velocity dispersion) of a ``toy'' emission line which occurs due to the gas motions in the regions marked in the upper panels.\label{fig:sloshing_vels}}
\end{center}
\end{figure}

Cold fronts are thus observable indicators of bulk motions related to large-scale structure growth and in general to all motions on spatially resolved scales in a medium with a substantial entropy gradient. Simulations have shown that characteristics of the observed velocity field depend strongly on the line of sight and its orientation with respect to the main direction of the cold front motion. For example, \citet{zuh2016} show that, at one extreme, if viewing the cluster along a line of sight perpendicular to the plane of its motion, both turbulence \textit{and} oppositely directed large-scale bulk motions contribute mainly to an increase in the observed velocity \textit{dispersion} ($\sim$100--200~km~s$^{-1}$) with little effect on the mean velocity shifts of the ICM. With this orientation, the morphology of the surface brightness edge associated with the cold front also resembles most closely a spiral pattern. At the other extreme, if the cold front is viewed from a line of sight within or close to the plane of its motion, the oppositely directed parts of its overall motion will appear as a mean velocity gradient across a large region, in addition to significant line broadening that is again the result of both random turbulent motions and the smoothly varying bulk motion (see Figure \ref{fig:sloshing_vels}). In this case, the spiral pattern in the surface brightness morphology disappears and is replaced by opposite and staggered, seemingly disconnected surface brightness edges \citep[see also similar conclusions in][]{Roediger2011}. At intermediate viewing angles between these two extremes, both mean velocity gradients as well as velocity dispersions will be observed, along with a weaker spiral pattern in the density distribution. Combined constraints on the morphology and kinematics of cold fronts therefore provide a powerful tool promising to unlock details about the microphysical properties that govern the dynamics of the ICM.  

\subsection{Subhalo infall and ram-pressure stripped tails}\label{sec:tails}
 
\begin{figure}
\begin{center}
\includegraphics[width=0.48\textwidth]{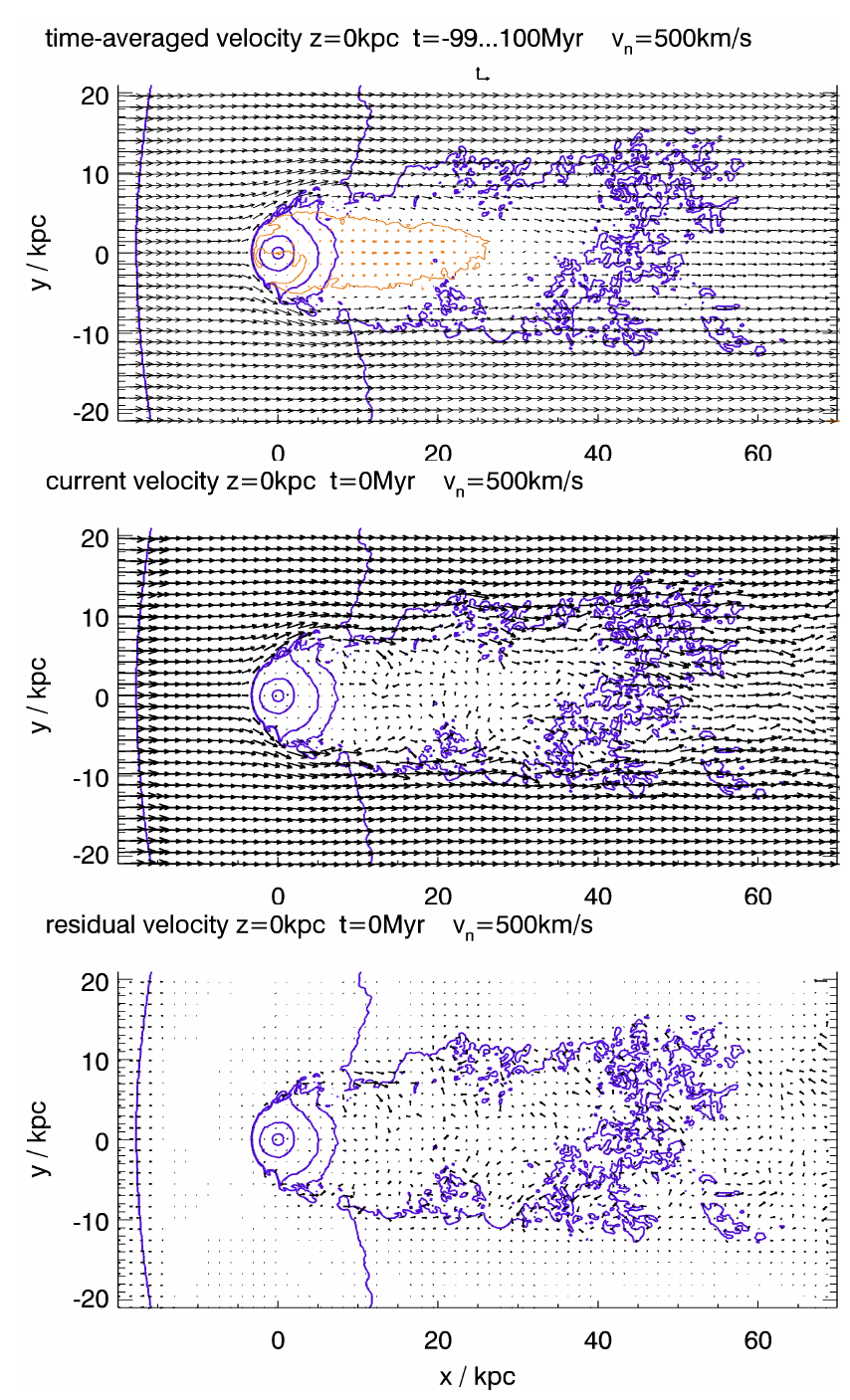}
\caption{Flow patterns around an elliptical galaxy falling into its host cluster, from \citet{Roediger15}.   The flow field is shown in the rest frame of the galaxy. The top, middle, and bottom panels show the time-average flow field, the current flow field, and the residual flow with respect to the averaged flow field, respectively. The purple contours show the gas density in a slice through the galaxy (logarithmic spacing, factor $\sqrt{10}$). The ICM is seen to flow around the remnant atmosphere similar to the flow around a blunt body. The downstream edge of the remnant atmosphere is at about the 4th innermost contour. Immediately downstream of the remnant atmosphere is the deadwater region, where on average the flow is directed back to the galaxy, i.e., into the upstream direction, as indicated by the red contour and the red flow vectors in the top panel. The deadwater region as well as the further downstream wake are turbulent (in the absence of mechanisms suppressing turbulence) for many atmosphere diameters downstream of the galaxy.}
\label{fig:flowpatterns}
\end{center}
\end{figure}

Another obvious location to look for ICM motions and turbulence are the tails and wakes of subclusters or galaxies falling into their host clusters. Well-known examples include ESO 137-001 in Abell 3627 \citep{sun10}, elliptical galaxies in the Virgo cluster (M86, M89, M60, M49; described in, e.g., \citealt{Ehlert2013,Kraft2017,Wood2017,Kraft2011}), and the elliptical galaxy NGC 1404 in Fornax \citep[e.g.][]{Su2017a,Sheardown2018}. More recently, several apparently ram pressure stripped groups or galaxies with very long tails (several 100 kpc) have been discovered in the outskirts of massive clusters, e.g, in A2142, Hydra~A or A85 (\citealt{eckert14,ichinohe2015,eckert17b,DeGrandi16}). The previously known case of the NGC 4839 group in the Coma cluster \citep{Neumann2001,Lyskova2018} belongs into the same category.

In the purely hydrodynamical case, the ICM flow patterns in and around such stripped subclusters or galaxies depend on the stage of the cluster crossing. The simplest case is the well-developed infall phase, where flow patterns of the ambient ICM closely follow the flow around a blunt body (\citealt{Roediger15}, see Fig.\ref{fig:flowpatterns}). Along the upstream edge, the pressure follows the well-known distribution of an enhanced pressure at the upstream stagnation point, decreasing towards the sides of the stripped atmosphere. Observational data now have reached a quality where these pressure gradients along the upstream edge can be measured and used to infer the 3D motion of the subcluster, as done by \citet{Su2017a} for NGC 1404 in Fornax. Downstream of the stripped atmosphere, the flow is characterised by a downstream deadwater region and a turbulent wake.

If the ICM can indeed be modelled largely by pure hydrodynamics, such wakes are potentially very long-lived, similar to the wake of a bullet in air shown in \citet{vanDyke1982}. A notable difference between the flow past a blunt body and a gas-stripped subcluster is the fact that the shape of the ``body'' in the flow changes. Specifically, the subcluster's gas is stripped predominantly in the upstream region and along its side, but the downstream atmosphere is shielded from the ICM head wind and can persist as a bright tail even after the subcluster passes the core of the main cluster. At this stage, the ICM flow patterns become more complex, as the subcluster moves towards the apocentre of its orbit. The rapid decrease of ram pressure experienced by the subcluster leads to its tail falling back towards the subcluster potential and overshooting the potential centre in a slingshot fashion. Likely several of the very long tails of subclusters in the main cluster outskirts are such slingshot tails and not simple ram pressure tails (as proposed, for example, for Abell 168, \citealt{Hallman2004}, and the Pandora Cluster, \citealt{Owers2011,Merten2011}). While the subcluster is lingering near its apocentre, the ICM flow patterns are far from the regular blunt-body-in-a-flow, and observations need to be interpreted carefully (for details see Sheardown et al, in submitted).
Beyond the pure hydrodynamic view, \citet{Vijayaraghavan2017,Vijayaraghavan2017a} show that unsuppressed thermal conduction, even saturated, would erase the tails of stripped ellipticals, while anisotropic suppression of thermal conduction due to magnetic fields allows the tails to survive.

\subsection{AGN feedback} 
\label{sec:AGNfeedback} 

While mergers are certainly the main source of gas motions in the bulk of the ICM, in the cores of relaxed clusters the energy input from the central supermassive black holes is plausibly the dominant mechanism that perturbs the gas. Here, the ICM radiative cooling time $t_{cool}$ is typically an order of magnitude (or more) shorter than the Hubble time, implying that without an external source of energy the gas should flow from radius $r$ towards the center of the potential well with the velocity $v\sim \frac{r}{t_{cool}}$ \citep[see][for a review of the cooling flow paradigm]{Fabian94} and accumulate there in the form of stars or cold gas. However, such a concept contradicts many observational constraints \citep[for a review, see][]{peterson2006}, suggesting that there must be a source of energy that compensates for the ICM radiative losses.

It is well established that most galaxies harbor a central SMBH with masses ranging between $10^6-10^{10}\,M_\odot$ (e.g., \citealt{kormendy13} for a review). The amount of energy released by such black holes is sufficient to expel the gas from galaxy-size halos \citep[e.g.][]{silk_rees98} or compensate gas cooling losses in more massive systems \citep[e.g.][see however \citealt{fujita_reiprich_2004}]{ciotti_ostriker_2001}. 
Indeed, the total feedback energy related to such SMBHs is $\varepsilon_{\rm m} M_{\rm BH}c^2\approx 10^{58}-10^{62}$ erg (with a mechanical efficiency $\varepsilon_{\rm m} =3\%$; \citealt{sadowski2017}).
Early X-ray and radio observations revealed that in many cool-core clusters there are bubbles of radio-bright plasma \citep[e.g.][]{Boehringer93} associated with the central SMBHs, which apparently are able to inflate cavities (X-ray dim regions). Such cavities are expected to be buoyant and must rise in the stratified atmospheres \citep{1973Natur.244...80G}. The comparison of the inflation and buoyancy time scales for the cavities led to the conclusion that the characteristic mechanical powers of the SMBHs in the Perseus \citep{Churazov00} and Hydra A \citep{McNamara00} clusters are comparable with the ICM cooling losses. This conclusion has been confirmed with \emph{Chandra} and \emph{XMM-Newton} observations for a large sample of clusters and groups spanning a wide range of masses and luminosities \citep[e.g.][see \citet{Fabian12,McNamara12} for reviews]{Birzan04,Hlavacek-Larrondo12}, suggesting that feedback from AGNs can prevent catastrophic cooling of the gas.  Here we focus on the gas velocities induced by AGN feedback in conditions relevant for galaxy clusters, rather than for lower mass systems, where different aspects of AGN-gas interactions might be important. 

By now, there is a rich landscape of models that consider different flavors of the AGN feedback \citep[see, e.g., the related chapter in this book,][]{WernerISSI}. While there is a general agreement on the importance of AGN feedback for the energy balance in cluster cores, the uncertainties in the form of the energy release by the AGN and in the physical mechanisms that lead to the energy dissipation are still very large. Measuring gas velocities can eliminate some of these uncertainties. 

The first group of models assumes that bubbles are filled with relativistic particles or very hot plasma, which remain confined within the bubbles, implying that there is no direct heat exchange with the ICM. In these models, the energy goes from the bubbles to the ICM via, e.g., the generation of sound or internal waves, turbulence in the wake of the bubbles, and entrainment of the low entropy gas, which all have different signatures in the observed gas velocities. 

A powerful injection of energy into an unperturbed hydrostatic atmosphere is naturally accompanied by shocks \citep[e.g.][]{Heinz98,Reynolds02}. This is true for spherical bubbles during their initial (fast) expansion phase and even more important for collimated outflows with large momenta. Many examples of weak shocks/sound waves are found in cluster cores \citep[e.g.][]{Fabian03,McNamara05,For07, Forman17,Randall15}. This led to the suggestion that most of the AGN energy goes into the generation of shocks and sound waves, which propagate radially and eventually dissipate due to the viscosity and conductivity of the ICM \citep[e.g.][]{Fabian03,Fabian06,fabian2017,Ruszkowski04,gaspari2011,barai16,Li17}. Other studies argue that the fraction of energy that is carried by sound waves is subdominant \citep[e.g.][]{Churazov00,Zhu16,Forman17,Tang_Churazov17} since, for a very fast expansion velocity, the energy is dissipated locally by a strong shock, while for a slow expansion the generation of sound waves is inherently inefficient. Nevertheless, if the outflow is collimated and has significant momentum, so that the head of the jet is always moving trans-sonically, sound waves can still play an important role. In terms of gas velocities, the sound waves scenario predicts quasi-spherical and almost radial motions (at some distance from the cluster center). The wave amplitudes decrease with radius due to the spherical geometrical factor and due to dissipation. In this scenario, the observed X-ray lines would be very broad and have boxy shapes towards the center (for spatially resolved wave patterns, but might be peaked for unresolved ones) and become progressively narrower with the projected distance from the center \citep[e.g.][]{rebusco2008,heinz10,zhuravleva2011}.  

A quasi-continuous energy injection scenario argues instead that the dominant fraction of energy goes into the enthalpy of the bubbles \citep[][]{Churazov00}. While there is a generic energy conservation argument stating that the buoyantly rising bubbles transfer most of their energy to the gas after crossing several pressure scale heights of the atmosphere \citep[][]{Churazov01,Begelman01,Churazov02}, this argument does not specify how the energy is extracted from the bubbles and how this energy is dissipated. Depending on the properties of the ICM (e.g., viscosity) and characteristics of the bubbles, the energy can be converted, for instance, into turbulence in the wake of the bubble or into internal waves \citep[e.g.][]{Churazov00,Omma_Binney04,gaspari2012b,Zhang18}. The latter scenario is attractive since the internal waves are trapped in cluster cores \citep[][]{Balbus_Soker90}, but can spread the energy in the tangential direction. As with sound waves, several other studies \citep[e.g.][]{Reynolds15,Bambic18} argue that internal waves are not playing an important role in the energy flows in cluster cores (see, e.g., the mixing scenarios below), or that the waves do not propagate fast enough in the vertical direction. The latter problem is alleviated by noting that the radial energy transport is provided by the bubbles themselves rather than by the waves. In terms of the velocity field characteristics, the main difference with the sound waves scenario is the predominance of the tangential over radial velocities. In contrast to radial sound waves, the lines could be narrower towards the cluster center but become broader at larger radii \cite[e.g.][]{rebusco2008,zhuravleva2011}. This conclusion may also depend on whether the spatial pattern is resolved or whether one observes the line from the entire region within a given radius. If the bubbles rise fast enough, the excitation of internal waves is inefficient and a hydrodynamic drag related to the turbulence in the wake of the bubble may dominate. 

The energy (enthalpy) of the bubbles can also go into entrainment of the low entropy gas from the central region to larger radii. Examples of structures that are likely formed by such entrainment are found in X-ray observations of many clusters \citep[e.g.][]{Simionescu08,Simionescu09,Gitti11,kirkpatrick2009,kirkpatrick2011,kirkpatrick2015}. Thermal instabilities are likely to develop in such entrained flows \citep[as pointed out by][]{Churazov01}. Indeed, cool core clusters are frequently observed to host filamentary nebulae of multi-phase gas with temperatures ranging from $<$100 — 10$^{4}$ K (see Section \ref{sect_optical}). 

\begin{figure}
\hspace{-0.3cm}
\includegraphics[width=0.51\textwidth]{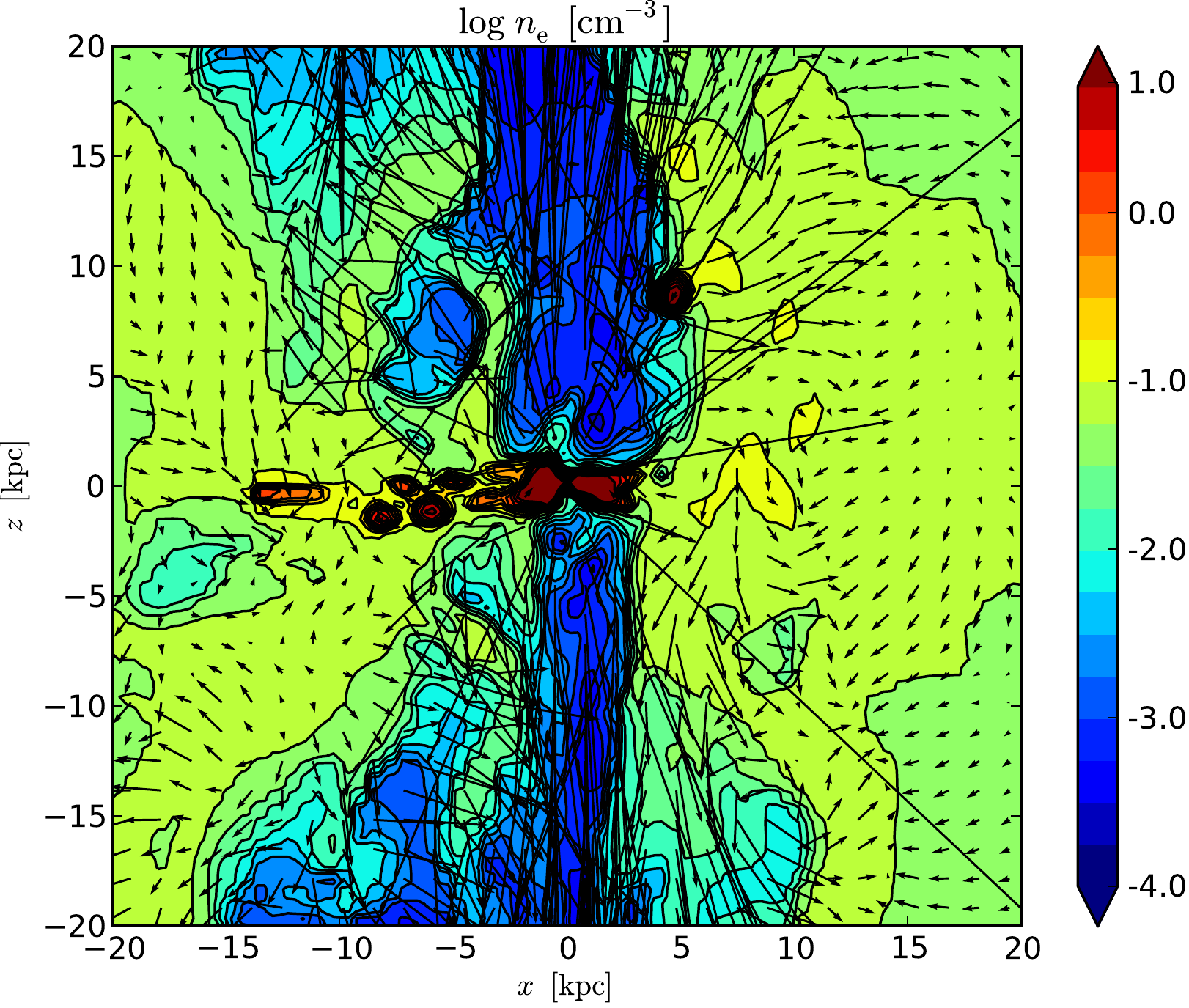}
\includegraphics[width=0.51\textwidth]{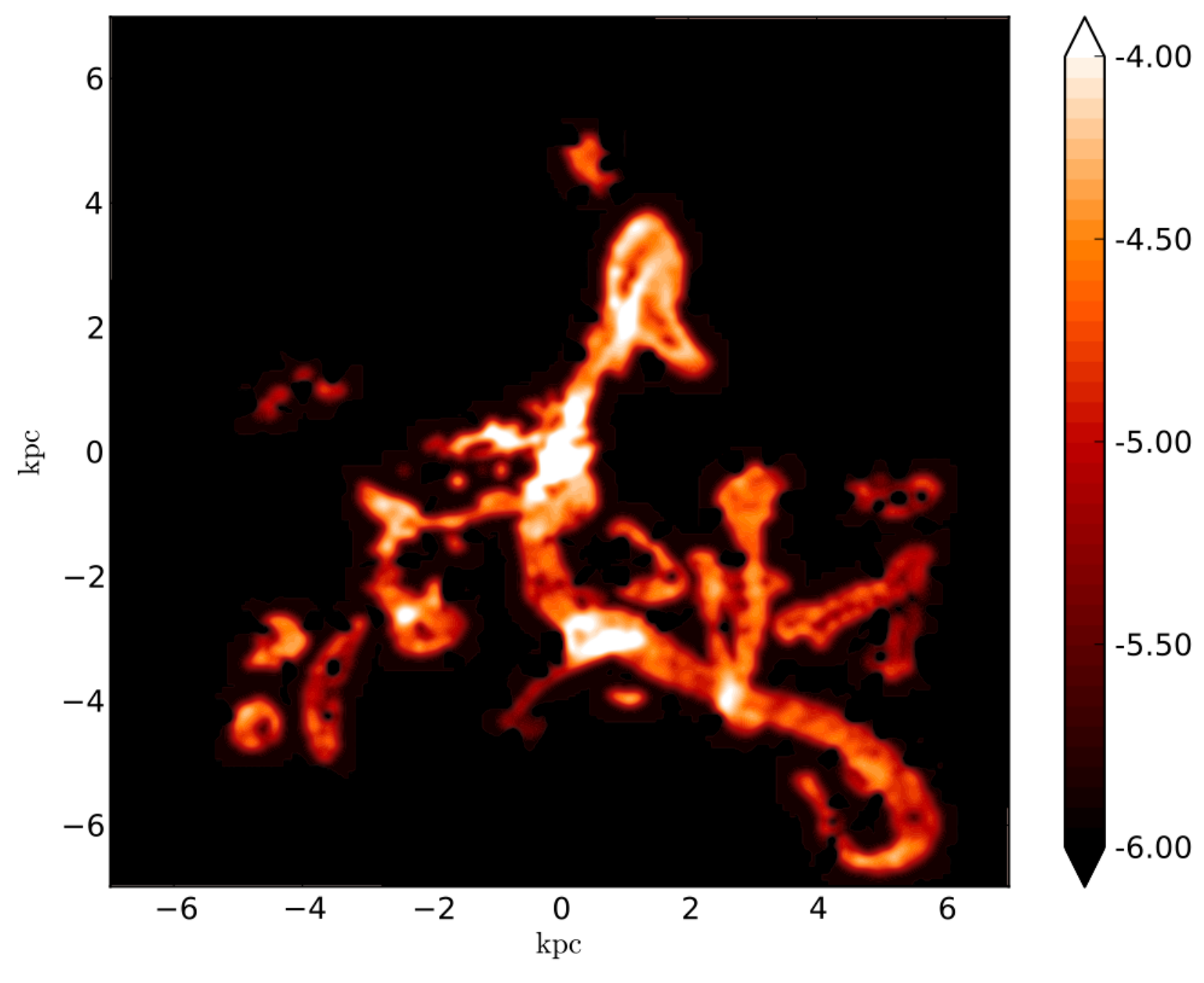}
\caption{The tight relation between turbulence and AGN feedback and feeding properties in X-ray halos. Left: electron number density cross-section of a BCG, showing an ultrafast outflow/jet triggered by a previous CCA rain phase (clouds in red) which generates large-scale cavities, shocks, and chaotic motions (velocity field overlaid as black arrows); reproduced from \citet{gaspari13rev}. Right: surface brightness map of the condensed gas -- turbulence-driven density fluctuations have become nonlinear, leading to a top-down multiphase condensation phase of warm ($\sim$\,$10^4$ K) filaments and clouds out of the diffuse hot plasma, which will soon boost the SMBH accretion rate via CCA and trigger a new generation of AGN jets/outflows; reproduced from \citet{gaspari2017}.
}
\label{fig:AGN_turb}
\end{figure}

The formation and evolution of these filaments has been studied in detail through a suite of numerical simulations that include the effect of radiative cooling. 
The cooling instability has been discussed intensively in the cooling flow models \citep[e.g.][]{fabian_nulsen77,mathews78}, although \cite{malagoli1987,balbus1989} have shown that linear perturbations do not grow.  \citet{Pizzolatto_Soker05} suggested that \textit{non-linear} density perturbations, possibly seeded by an earlier radio jet, would cool to low temperatures (T$< 10^4$ K) and feed the black hole. 
In terms of the evolution of density perturbations evolving from either linearly idealized \citep{McCourt2012, Sharma2012} or realistic AGN jet conditions \citep[][see left panel in Fig. \ref{fig:AGN_turb}]{gaspari2012a, gaspari13rev}, it has been suggested that thermal instabilities can in fact develop throughout the volume of the cool core when the ratio of the cooling time and free-fall time, $t_{cool}/t_{ff}$, falls below roughly a factor of 10.
Numerical simulations went on to demonstrate that uplift of low-entropy ambient gas along the direction of the rising radio bubbles is at least as important to the production of cold clouds \citep{revaz2008,Li2014}. The conditions under which thermal instabilities develop are discussed in more detail in the chapter on ``Hot X-ray Atmospheres, Molecular Gas and AGN Feedback in Early Type Galaxies: A Topical Review'', in this volume \citep{WernerISSI}. In any case, it is generally clear that the combination of gas motions triggered by the AGN (either in the form of volume-filling turbulence or directed entrained outflows), together with a short central cooling time and consequently a low $t_{cool}/t_{ff}$, correlate with and are likely responsible for the onset of cooling of the X-ray emitting ICM and the formation of optical and sub-millimetre line emitting filaments.

The presence of thermal instabilities and cold(er) gas phases embedded in the ICM is thought to be crucial for regulating the AGN feeding and feedback cycle (Figure \ref{fig:AGN_turb}, right panel). 
The AGN jet feedback typically drives a turbulent velocity dispersion $\sigma_v \approx 100-300$ km\,s$^{-1}$ \citep[as shown by high-resolution hydrodynamical simulations in][]{gaspari2012a,gaspari2012b, gaspari2018, valentini15}, which enables both a self-consistent gentle circulation and the seeding of nonlinear (lognormal) gas fluctuations (see also Fig. \ref{fig:PSmulti}).
In analogy to Earth weather, warm extended filaments and cold clouds condense out of the hot ICM and rain toward the central SMBH \citep{gaspari2012a,gaspari15,gaspari2017,voit2015,Voit18,Prasad17}. Inelastic collisions in the central kpc region promote angular momentum cancellation and boost the AGN feeding rate by 100 fold, in a process known as Chaotic Cold Accretion (CCA; \citealt{gaspari13cca}; see also Sec.~\ref{sect_optical}). The accretion of cold gas triggers a new generation of AGN feedback injection. The resulting subsonic turbulence mixes the gas in an eddy turnover time, $t_{\rm eddy}=L/\sigma_v$, where L is the injection scale of the size of the bubble diameter. Later, as the cooling time lowers back again and $t_{\rm cool}/t_{\rm eddy}\approx 1$, a new cycle of CCA rain and self-regulated AGN feedback restarts (\citealt{GS17} for a review of the self-regulation). This duty cycle continues for several billion years and has to preserve the cool-core structure since at least redshift $z\sim 2$ (\citealt{McDonald17}). 

In addition, there are also many more numerical simulations that include other physical mechanisms, e.g., magnetic field effects \citep[see][]{Vernaleo06,Ruszkowski17a,Bambic18}, explicit viscosity \citep[][]{Reynolds15}, or streaming of cosmic rays \citep[][]{Bohringer_Morfill88,Loewenstein91,Wiener17,Ruszkowski17b}, whose distinct features are also partly reflected in the gas velocity fields.
The vast majority of models mentioned in this section predict mostly subsonic velocities in the bulk of the ICM within the cluster cooling radius\footnote{These arguments do not exclude though a possibility of having localized regions with supersonic motions, especially in the regions where the AGN-driven outflows/jets interact with the ICM \citep[e.g.][]{heinz10}.}. This is a rather natural result, given that a) strongly supersonic motions decay very quickly and are therefore short-lived and b) the large ratio between the cooling time and the free fall time implies that at least in a quasi-steady scenario supersonic motions in the bulk of the ICM are not required and would instead overheat the gas. Given the superposition between the motions due to AGN feedback and those driven by mergers on larger scales (as discussed in the previous subsections), the question arises whether the predictions by the large range of models described above are different enough to allow an easy differentiation between various scenarios.  

In a different group of models, the AGN injects hot (but not relativistic) plasma\footnote{Such plasma could also appear as the result of ICM heating by strong shocks, if the energy is released by an AGN in the form of short and powerful outbursts.} that can mix with the ICM \citep[][]{Brueggen_Kaiser02,Pizzolatto_Soker05,Reynolds15}. Such mixing is naturally present in numerical simulations that use grid-based schemes, but might be suppressed in SPH-based codes \citep[][]{Sijacki_springel06}. 
At low resolution, the numerical mixing might be relevant; however, in modern simulations with intermediate/high resolution, the mixing is dominated by physical turbulent mixing and hydrodynamical instabilities (such as Rayleigh-Taylor or Kelvin-Helmholtz). Mixing is a very efficient process of sharing the injected energy with large volumes of the ICM and other heating channels might be subdominant. 
CARMA maps show a deficit in the SZ signal coincident with the X-ray identified cavities in MS~0735.6+7421 \citep{Abdulla18}, suggesting that they are unlikely to be supported thermally, although extremely diffuse thermal plasma with temperatures in excess of hundreds of keV cannot be ruled out as the main component of the bubbles. If the injected plasma is very hot, but not relativistic, then the most important constraints on AGN heating might come not from the gas velocities but from the broad temperature distribution of the mixing plasmas.

\subsection{Other sources of turbulence in the intra-cluster medium}\label{sect_instab}

In addition to the structure formation process and AGN feedback, cluster member galaxies also influence the dynamics of the surrounding ICM, even in relaxed systems. \citet{ruszkowski2011} show that galaxies moving through the host cluster excite large-scale g-modes, leading to volume filling turbulence with a velocity dispersion in the central 100~kpc estimated at 150--200 km/s.

The level of gas motions in the ICM can also be affected by instabilities inherent to a stratified low-collisionality plasma in the presence of a weak magnetic field.
If anisotropic thermal conduction is the dominant mode of heat transport, instabilities can develop on macroscopic scales. Examples include the magnetothermal instability (MTI) \citep{balbus2000,parrish2005}, which is important in the case of a negative temperature gradient usually found in the cluster outskirts \citep{parrish12}, and the heat buoyancy instability (HBI) \citep{quataert2008}, which occurs for a positive temperature gradient (for instance in the cores of relaxed clusters, where the temperature increases as a function of radius). However, \citet{ruszkowski10} argued that even a very low level of turbulent perturbations, entirely consistent with the expectations for cosmological infall, galaxy motions, mergers, or AGN activity, can entirely alter the magnetic field distribution resulting at least from the HBI instability, preventing it from saturating.

Furthermore, any turbulent stresses and the resulting local changes in the magnetic field will trigger very fast micro-scale instabilities, such as the firehose, mirror, and gyrothermal instability \citep[e.g.][]{schekochihin2005,schekochihin2010,lyutikov2007}. These instabilities affect the large-scale transport properties of the ICM, including its effective viscosity and consequently its dynamics. \citet{kunz2011} propose that parallel viscous heating, due to the anisotropic damping of turbulent motions, is regulated by the saturation of micro-scale plasma instabilities; if this is assumed to be the source of heating that balances radiative cooling in the cores of galaxy clusters, the predicted ICM turbulent velocities are of order 100--200 km/s for an A1835-like cluster. 

\section{Observational probes of the ICM velocity}\label{sec:obs_probes}

\subsection{State of the art in the absence of high-resolution, nondispersive X-ray spectrometers}

\subsubsection{Line shifts}\label{line_shift}

The simplest manifestation of bulk motions in the ICM is the Doppler shift of X-ray spectral emission lines. The magnitude of the energy shift is approximately given by $\Delta E = E_{\rm line} \times v_{\rm bulk}/c$, where $E_{\rm line}$ is the expected position of the spectral line in the absence of gas motions, $v_{\rm bulk}$ is the line-of-sight velocity of the gas (note that motions in the plane of the sky cannot be detected in this way), and $c$ is the speed of light. Hence, these measurements are extremely challenging, since even a 1\% uncertainty in calibrating the detector gain corresponds to an error of 3000 km/s in the bulk velocity. Initial attempts to constrain the magnitude of gas motions using ASCA and later \emph{Chandra} and \emph{XMM-Newton} data had suggested large line-of-sight velocity gradients of order a few thousand km/s in the Perseus and Centaurus clusters \citep{dupke2001a,dupke2001b,dupke2006}, as well as the merging cluster Abell 576 \citep{dupke2007a}, although the associated uncertainties were themselves of order 1000 km/s. Using the $^{55}$Fe calibration source onboard the Suzaku satellite, the absolute energy scale calibration of the XIS detectors can be determined more accurately \citep[with a precision as good as 0.1\%,][]{ozawa2009}, allowing somewhat tighter constraints to be placed on the spectral line shifts from several galaxy clusters; at 90\% confidence, \citet{ota2007} place an upper limit of 1400 km/s line-of-sight velocity gradient in the Centaurus Cluster, and \citet{sugawara2009} find that the bulk motions in Abell 2319 do not exceed 2000 km/s. 

In addition to these upper limits, several detections of gas bulk motions were reported from Suzaku data. In the merging cluster Abell 2256, \citet{tamura2011} found a significant line of sight velocity gradient of 1500$\pm$300(statistical)$\pm$300 (systematic) km/s. In the Perseus Cluster, \citet{tamura2014} discovered a hint of gas bulk motions at the level of only -(150-300) km/s relative to NGC1275; this velocity gradient is spatially coincident with the cold-front located 45-90 kpc west of the cluster center. \citet{ota2016} performed a search for line shifts with respect to the expected rest frame in a sample of nearby clusters of galaxies with various X-ray morphologies observed with Suzaku; they report upper limits of 3000--4000 km/s in A 2199, A 2142, A 3667, and A 133, and hints of large bulk velocities in excess of the instrumental calibration uncertainty near the center of the cool-core cluster A2029 and in the subcluster of the merging cluster A2255.

In summary, X-ray CCD detectors typically indicate that gas bulk motions in cluster centers are below a few thousand km/s, but significant detections of line shifts remain elusive -- particularly in relaxed systems.

\subsubsection{Line broadening}\label{line_broad}

In addition to the centroids of X-ray spectral emission lines, their widths also hold important clues regarding the gas dynamics. Thermal motions of the respective ion, turbulent gas motions along the line of sight, and the response of the instrument all contribute in quadrature to the observed width of a line: $W_{\rm obs}^2=W_{\rm therm}^2+W_{\rm turb}^2+W_{\rm inst}^2$. The thermal motions can be calculated as 
$$W_{\rm therm}=\frac{\nu_0}{c}\sqrt{\frac{k_BT}{Am_p}}$$ 
where $k_B$ is the Boltzmann constant, $m_p$ the proton mass, $A$ the atomic
weight of the element, and $\nu_0$ the frequency of the transition. The turbulent line broadening meanwhile is written as
$$W_{\rm turb}=\frac{\nu_0}{c}\sigma_{v||}$$
where $\sigma_{v||}$ is the line of sight component of the turbulent velocity. For the case of isotropic turbulence, $\sigma_{v||}^2=\sigma_{v}^2/3$, where $\sigma_{v}$ is the root mean square of the three-dimensional turbulent velocity. Because only $\sigma_{v||}$ affects the observed spectral line widths, throughout this paper we define the observed turbulent velocities $v_{\rm turb}\equiv\sigma_{v||}$. 
It should be noted that both $W_{\rm therm}$ and $W_{\rm turb}$ are typically of order a few eV; hence, CCD spectrometers with $W_{\rm inst}>100$ eV cannot be used to constrain the level of turbulent motions in the ICM because the instrumental line broadening dominates over the expected Doppler broadening. 

The first successful attempts to constrain the level of turbulence in galaxy cluster cores from X-ray line widths were instead performed using the Reflection Grating Spectrometers (RGS) onboard \emph{XMM-Newton}. The RGS has a remarkable energy resolution of only $\sim2-3$ eV at 1~keV for a point-like source. However, its nature as a slit-less spectrometer complicates the analysis of extended objects such as diffuse, nearby clusters of galaxies.
This is because photons originating from a region that is offset in the direction along the dispersion axis are slightly shifted in wavelength, resulting in changes in the line position and width. This effect can be corrected for, approximately, by using a surface brightness profile extracted along the dispersion direction and convolving it with the instrumental line shape. Note, however, that the spatial distribution of various ions (O, Fe, etc) may not be exactly the same as that of the average gas density (for example, due to a radial gradient of the ICM metallicity), making this approach somewhat uncertain. 
It is clear that the correction to the line width is smallest (and hence uncertainties in this correction have the least effect on the measurements) for clusters with a very compact, sharp central X-ray emission peak. Taking advantage of this strategy, \citet{sanders2010} placed the first direct limit on the turbulent velocity in Abell 1835 of $v_{\rm turb}<274$ km/s (at 90\% confidence), while \citet{bulbul2012} report an upper limit of $v_{\rm turb}<206$ km/s in Abell 3112. \citet{sanders2013} and \citet{pinto2015} expanded this method to larger samples of clusters, groups of galaxies, and elliptical galaxies. The results by \citet{pinto2015} were based on the CHEmical Enrichment RGS Sample (CHEERS), designed as a ``complete'' sample of high-quality RGS cluster spectra within $z<0.1$ that can be obtained within a reasonable exposure time of less than 200 ks each \citep{dePlaa2017}. This study found that one half of the 44 objects in the sample had turbulent velocities $v_{\rm turb}\equiv\sigma_{v||}$ below 350 km/s (at 90\% confidence), while in about a quarter of the objects values as high as $v_{\rm turb}>700$ km/s were still allowed by the data.


\subsubsection{Resonant scattering}\label{res_scat}
Although the ICM is generally optically thin, if the gas turbulent velocities are negligible compared to the thermal velocities of ions, then several of the brightest X-ray emission lines are expected to become moderately optically thick in the cores of galaxy clusters, groups and massive elliptical galaxies. Because the optical depth is highest in the central region of a cluster, a fraction of the photons originating from that region are expected to be resonantly scattered away from the line-of-sight, causing an apparent suppression of the resonant line relative to other lines \citep{gilfanov1987}. Since the optical depth is sensitive to the turbulent broadening, one can probe gas velocities in the hot haloes of clusters and galaxies by measuring the apparent flux suppression, even if the lines are not fully resolved spectrally. We refer the reader to the chapter on ``X-ray spectroscopy of galaxy clusters: beyond the CIE modeling'' in this volume \citep{GuISSI} for a more detailed discussion on the effects of resonant scattering on the X-ray spectra of the ICM; here, we focus solely on the turbulent velocity constraints obtained from this method.

Early \emph{XMM-Newton} CCD spectrometer data did not reveal a strong suppression of the resonant He-like Fe line at 6.7 keV in the core of the Perseus Cluster \citep{2004MNRAS.347...29C,2004ApJ...600..670G}, which at face value would imply that strong turbulent motions are present in this region. It was later shown that, with the limited energy resolution of CCDs, it is difficult to cleanly single out the magnitude of resonant scattering when variations of gas temperature and metallicity are present \citep{2013MNRAS.435.3111Z}. This difficulty also explains some controversial results obtained for other clusters \citep[e.g.][]{1999ApJ...519L.119K,2000AdSpR..25..603A,2001ApJ...550L..31M,2002A&A...391..903S,2006MNRAS.370...63S}. 

\begin{figure*}
\begin{minipage}{\textwidth}
\includegraphics[width=\textwidth]{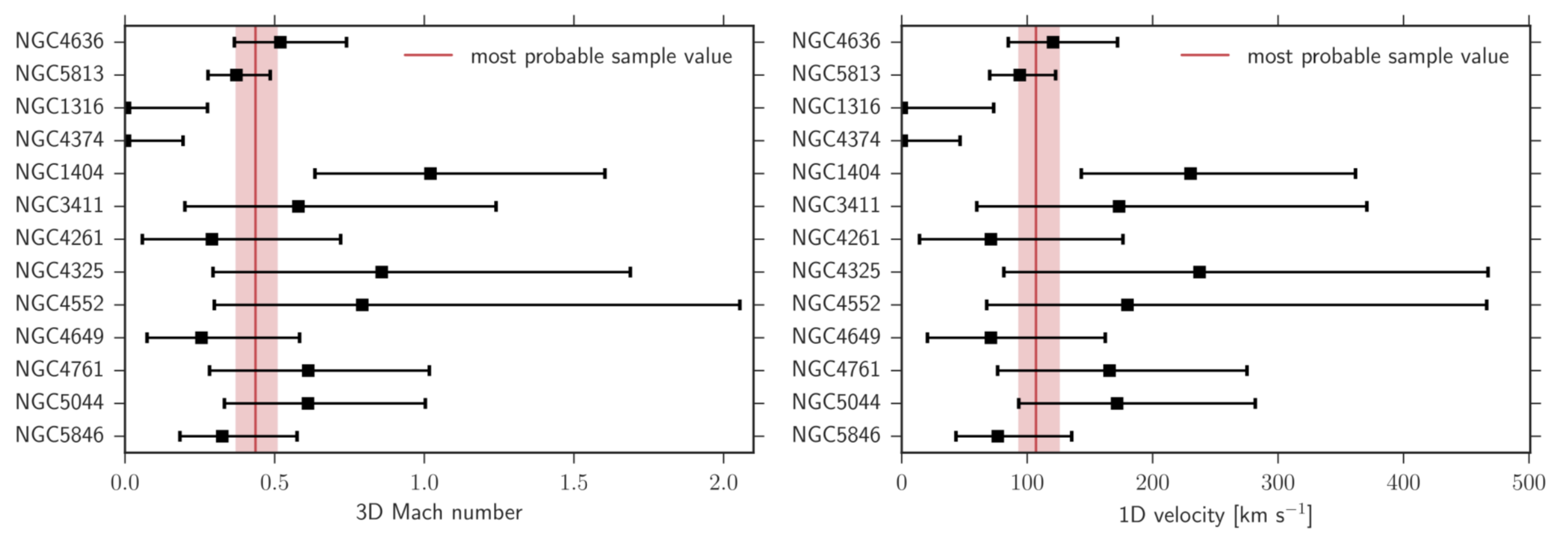}
\end{minipage}
\caption{Combined resonant scattering and line broadening constraints on 3D Mach number (left) and 1D turbulent velocity amplitude (right) for 13 galaxies with 1$\sigma$ uncertainties. Red lines and regions show best fit most probable sample values. Adapted from \citet{ogorzalek17}. 
\label{fig:ogorz_vel}
}
\end{figure*}

Clear evidence of the resonant scattering effect was found using high-resolution \emph{XMM-Newton} RGS spectra of the cooler gas in the cores of giant elliptical galaxies and groups. A strong suppression of the Ne-like Fe line at 15.01\,\AA\, was detected in many galaxies, including NGC4636, NGC1404, NGC5813 and NGC4472 \citep{2002ApJ...579..600X,2003ASPC..301...23K,2009MNRAS.398...23W,deplaa12}. \citet{2009MNRAS.398...23W} and \citet{deplaa12} modeled the resonant scattering effect for different values of the characteristic turbulent velocities and, by comparing the data to the models, found $v_{\rm turb}<100$ km/s in the inner kpc region in NGC 4636, $320<v_{\rm turb}<720$ km/s in NGC5044 and $140<v_{\rm turb}<540$ km/s in NGC5813. More recently, \citet{ogorzalek17} extended the resonant scattering analysis to a sample of 13 elliptical galaxies, finding velocities ranging between 0--120 km/s in the cores of four massive galaxies and placing lower limits of 20--100 km/s in the remaining nine objects. Combining the line broadening results of \citet{pinto2015} and the resonant scattering results, \citet{ogorzalek17} obtained velocity constraints for all 13 galaxies, with the sample best-fit mean value of $v_{\rm turb}\sim$ 110 km/s, 3D Mach number of $\sim$ 0.45, and non-thermal pressure fraction $\sim$ 6 per cent (Fig. \ref{fig:ogorz_vel}). 

\subsubsection{Surface brightness fluctuations and power spectra}\label{sect_fluct} 

The idea behind the gas velocity measurements from surface brightness fluctuations is simple: any deviations of the hot gas from hydrostatic equilibrium should be associated with gas motions leading to perturbations in X-ray images relative to a smooth, undisturbed model. 
Various physical mechanisms behind this correlation as well as different approaches to extract this information from observational data have been discussed in \citet{Sch04,Chu12,San12,Gas13,Gas14,Zhu14b,Zhu14a}. The study of \citet{Sch04} was focused on the case of isotropic turbulence in a homogeneous fluid that leads to pressure fluctuations scaling as the square of gas velocities. In \citet{Chu12} several other processes contributing to X-ray surface brightness fluctuations have been added, including perturbations of gravitational potential, entropy and density perturbations due to turbulent motions in stratified atmospheres, density variations in sound waves, metallicity variations and bubbles of relativistic plasma.  

\begin{figure}
\hspace{-0.3cm}
\centering
\includegraphics[width=0.499\textwidth]{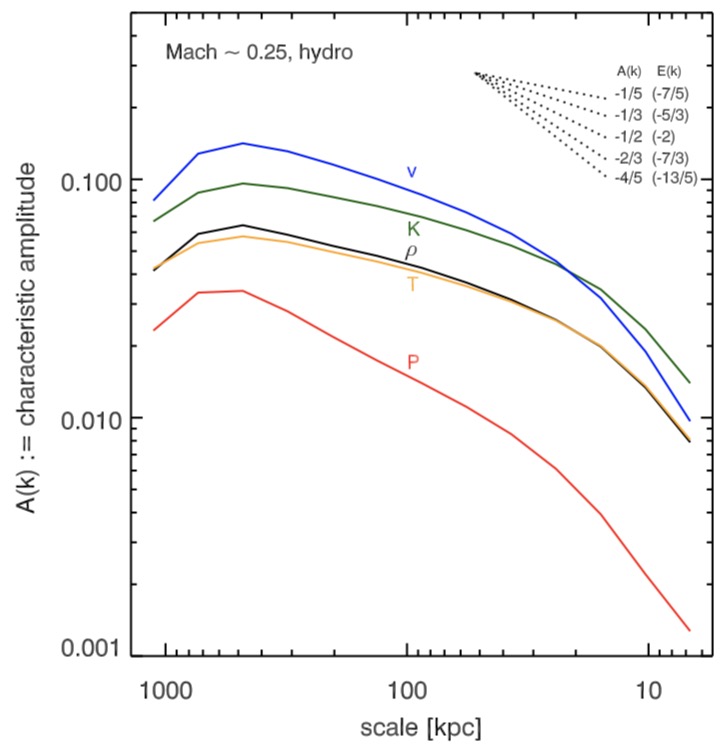}
\includegraphics[width=0.499\textwidth]{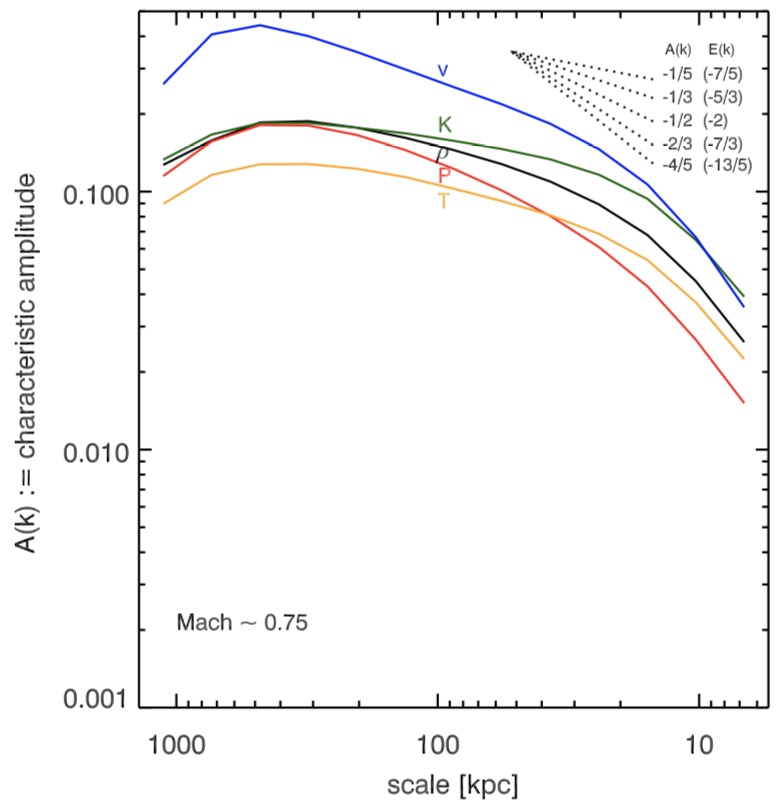}
\caption{Power spectra of all thermodynamic fluctuations (density, temperature, pressure, entropy) from high-resolution hydrodynamical simulations of turbulence in a stratified Coma-like cluster (no conduction). Left: Mach 0.25 (weak) turbulence. Right: Mach 0.75 (strong) turbulence. It is clear that the amplitude of thermodynamic fluctuations is linearly related to the turbulent Mach number, while the slope depends on the specific transport properties of the gas (and does not necessarily behave as a passive scalar). Reproduced from \citet{Gas14}.
}
\label{fig:PSmulti}
\end{figure}

By using high-resolution 3D hydrodynamic simulations probing varying levels of turbulence in the electron-ion ICM, \citet{Gas13} first showed that the normalization (amplitude $A_\delta$) of the X-ray density power spectrum is {\it linearly} related to the turbulent Mach number (unlike the above suggestions of a square dependence): $A_\delta \simeq 0.25\,{\rm Mach_{\rm 3D}}$. 
The same suite of simulations showed that it is possible to convert the density spectrum into a velocity spectrum (Fig.~\ref{fig:PSmulti}), although below the injection scale, the linear relation tends to progressively loosen up, in particular in the presence of strong thermal conduction. The steepening of the slope (from Kolmogorov toward Burgers) represents another way to probe the plasma microphysics in the ICM. 
\citet{Gas14} further showed the power spectra evolution of all the thermodynamic quantities: for $M < 0.5$ the entropy power spectrum (internal waves) dominates over that of pressure (sound waves), while for $M > 0.5$ pressure waves start to become substantial. Increasing levels of turbulent motions thus move the ICM from the isobaric to the more compressive, adiabatic regime (Fig.~\ref{fig:PSmulti}).

When all perturbations in a stably stratified atmosphere are infinitesimal and all motions are slow (${\rm M} \ll 1$) and the Boussinesq approximation is used, there is a linear relation between the velocity and density fluctuations, corresponding to internal waves. These density perturbations are sourced by the entropy gradient present in the system. \citet{Zhu14b} showed that the linear relation may extend to the non-linear regime, such that $\left ( \delta \rho_k /\rho\right)^2 \approx \eta^2 \left ( v_{k}/c_s\right)^2$, where $\delta \rho_k$ and $v_{k}$ are some suitably defined fluctuation amplitudes at spatial scales $k^{-1}$ and $\eta\sim 1$ when the one dimensional velocity component is considered. Moreover, as the turbulent cascade proceeds to smaller scales, the role of buoyancy becomes negligible below the so-called Ozmidov scale. On these scales and under the above approximations, the entropy behaves as a passive scalar advected by the velocity field and, according to the Obukhov-Corrsin theory, the velocity and the scalar power spectra should have similar slopes, while the relative normalization is set by the coefficient $\eta$. Tests using relaxed clusters from cosmological simulations show $\eta\approx 1$, albeit with a 30\% scatter. 

Observationally, there are a few main hurdles associated with the X-ray fluctuation power spectral analysis. First, the choice of ``unperturbed'' model is not unique, and there is no perfect scale separation between the measured perturbations and the model. Thereby, the amplitude of density fluctuations measured on large scales may be affected by the choice of unperturbed model.
Second, the final dynamic range of the observed power spectra is usually no larger than 10\,--\,30. Third, thermodynamic variables in full ICM hydrodynamics are not perfect scalars, thus at small scales (large $k$ modes) deviations from the approximate linear conversion are expected (Fig.\,\ref{fig:PSmulti}).
Keeping in mind such uncertainties and caveats, using the above relation, velocity power spectra have been retrieved from the observed density power spectra in the cool cores of several galaxy clusters.

\begin{figure*}
\includegraphics[width=1\textwidth]{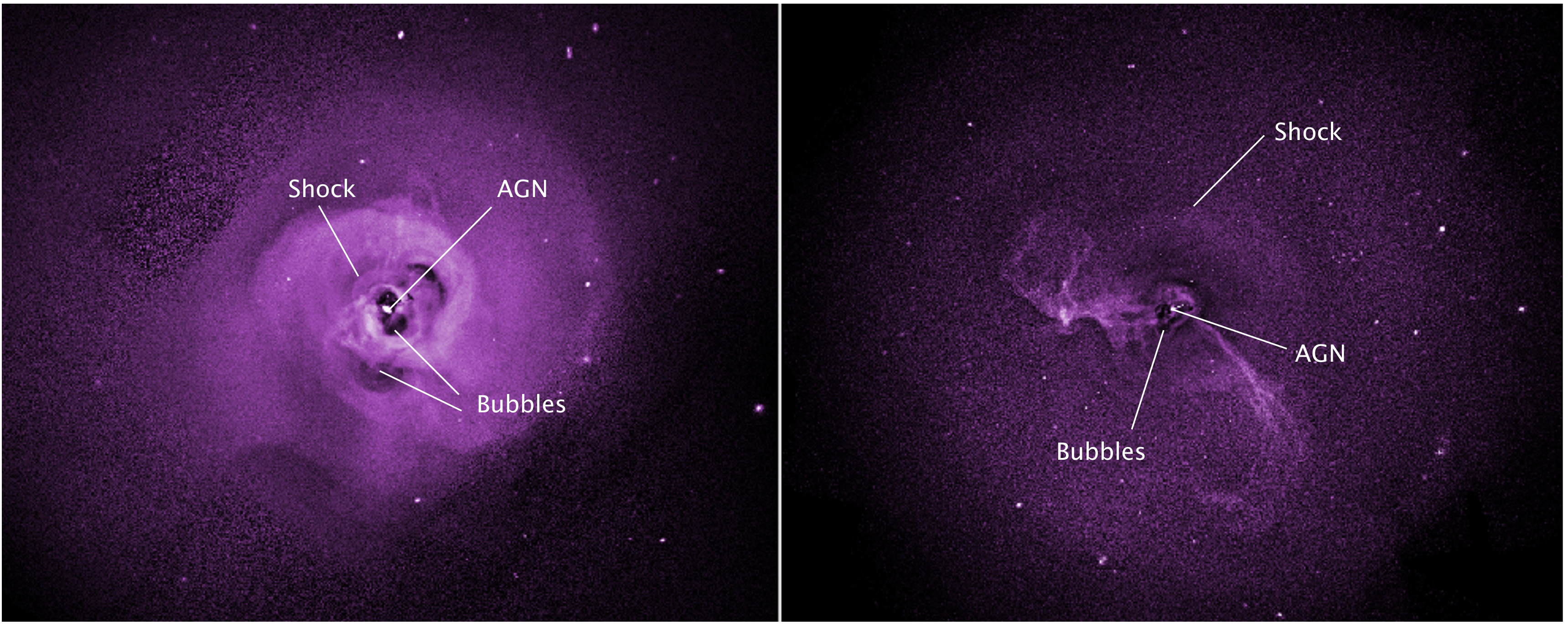}
\caption{\emph{Chandra} residual images of the cool cores in the Perseus (left) and Virgo (right) galaxy clusters. Adapted from NASA/CXC/Stanford/Zhuravleva et al. 
\label{fig:pers_virgo}
}
\end{figure*}

\begin{figure*}
\begin{center}
\includegraphics[width=0.85\textwidth]{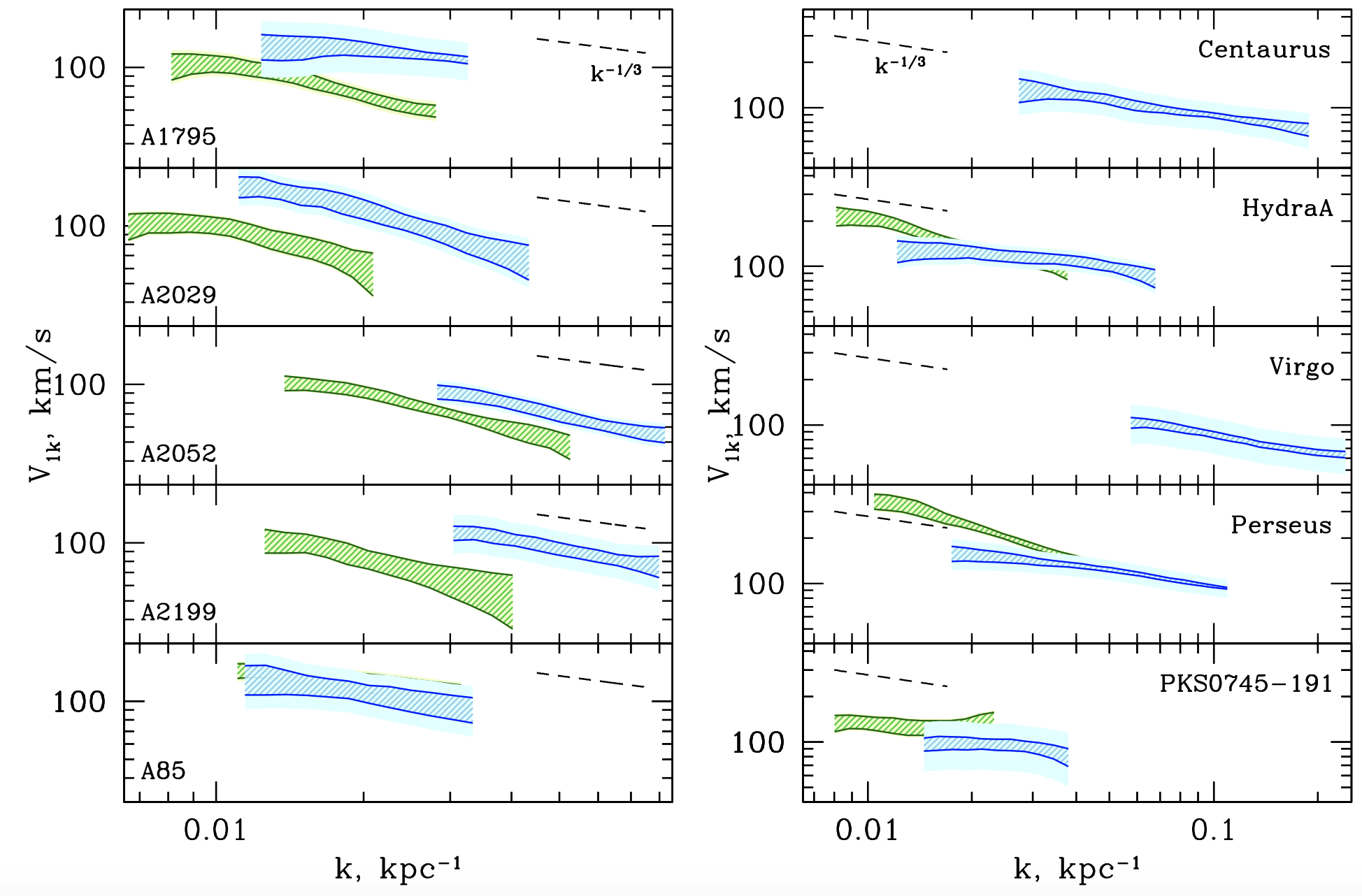}
\end{center}
\caption{Amplitude of the one-component velocity of gas motions versus wavenumber $k = 1/l$ determined from the analysis of X-ray surface brightness fluctuation power spectra for a sample of galaxy clusters \citep[from][]{Zhu17}.
Blue: velocities measured in the inner, half-cool-core regions. Green: same in the outer annulus of the cool core. Dashed lines show the
slope of the amplitude for pure Kolmogorov turbulence (with arbitrary normalization). Hatched regions: velocities calculated using the
mean sound speed in the gas. Solid regions: the spread of the velocity if maximal and minimal values of the sound speed are used.}
\label{fig:sb_vel}
\end{figure*}

Deep \emph{Chandra} observations of Perseus and Virgo show that the hot gas in the core regions of these clusters is disturbed (Fig. \ref{fig:pers_virgo}) most likely by the activity of a powerful AGN \citep{For07, Fab11} residing in the center of both systems. This suggests that the gas may be turbulent. \citet{Zhu14a} used the power spectra of gas density fluctuations in these two galaxy cluster cores to report the first constraints on the corresponding velocity power spectra (for details of the analysis, uncertainties, and spectra in all regions in Perseus see \citealt{Zhu15}). Later work confirmed that about 80\% (50\%) of the total variance of perturbations in the Perseus (Virgo) core have isobaric nature, i.e. are consistent with subsonic gas motions in pressure equilibrium and/or gas cooling \citep{Are16,Zhu16,Chu16}. \citet{Wal15} measured velocity power spectra also in the core of the Centaurus cluster. The velocities of gas motions inferred in this way vary between $\sim$ 80 and $\sim$ 160 km/s on spatial scales of $\sim 10-30$ kpc in the Perseus core, between $\sim 40$ and $\sim 80$ km/s on $\sim 2-10$ kpc scales in the Virgo core, and between $\sim 100$ and $\sim 150$ km/s on scales of $\sim 4-10$ kpc in Centaurus. Recently, velocity power spectra have been measured based on X-ray surface brightness fluctuations in several other nearby, cool core clusters that have deep \emph{Chandra} observations \citep{Zhu17}. These constraints are shown in Figure \ref{fig:sb_vel}. The typical velocities are consistent with those in Perseus, namely, between $\sim 100$ and 150 km/s on scales $<$ 50 kpc, and could be up to $\sim 300$ km/s on scales $\sim 100$ kpc. 

\citet{Hof16} analyzed deep {\it Chandra} observations of 33 well-known clusters, focusing instead on fluctuations of the projected thermodynamic quantities (the so-called pseudo density, temperature, pressure, entropy) within the core and outskirt
($>$100 kpc) regions. They argue that the observed fluctuations correspond to a sample averaged 1D Mach number of $0.16\pm0.07$.

\citet{Kha16} presented a study of thermal Sunyaev Zel'dovich (SZ) fluctuations in a hot (Coma) cluster with {\it Planck}, thus providing constraints on the turbulent power at the 500 kpc injection scale (mainly associated with mergers) with retrieved 3D Mach number up to 0.8. The SZ power spectrum directly probes pressure fluctuations, which are thus complementary to the small-scale X-ray density fluctuations. Indeed, this study showed that the Coma X-ray power spectrum (with amplitude of $\sim$\,10\%; \citealt{Chu12,Gas13}) can be extended and joined with the large-scale SZ power spectrum (with amplitude of $>$\,30\%).
Moreover, the observed Coma (X-ray) power spectrum shows a slope consistent with the Kolmogorov value ($A_\delta \propto k^{-1/3}$), indicating a highly suppressed thermal conductivity ($f\approx 10^{-3}$). Other studies focused on measuring perturbations on large spatial scales via X-ray images have been carried out by \citet[]{Shi01,Fin05,Kaw08,Gu09}. 

This method has also been used to investigate the connection between turbulence and particle re-acceleration in radio haloes. 
\citet{Eck17a} combined for the first time the large-scale power spectrum amplitude of X-ray fluctuations with the 1.4\,GHz radio power in over 50 clusters. Remarkably, those clusters hosting giant radio halos follow the relation $P_{\rm radio}\propto \sigma_v^3$ defined by the turbulent energy flux rate. \citet{bonafede18} used the spectrum of density fluctuations from {\it Chandra} to suggest a different ratio of kinematic over thermal energy in the regions with and without extended radio halo in MACS J0717.5+3745, further strengthening the turbulent re-acceleration scenario. For a comprehensive review of the role of turbulence in magnetic field amplification and cosmic ray acceleration, we refer the reader to the chapters by \citet{DonnertISSI} and \citet{vanWeerenISSI}, included in this volume.

\begin{figure}[t]
\begin{center}
\includegraphics[width=\textwidth]{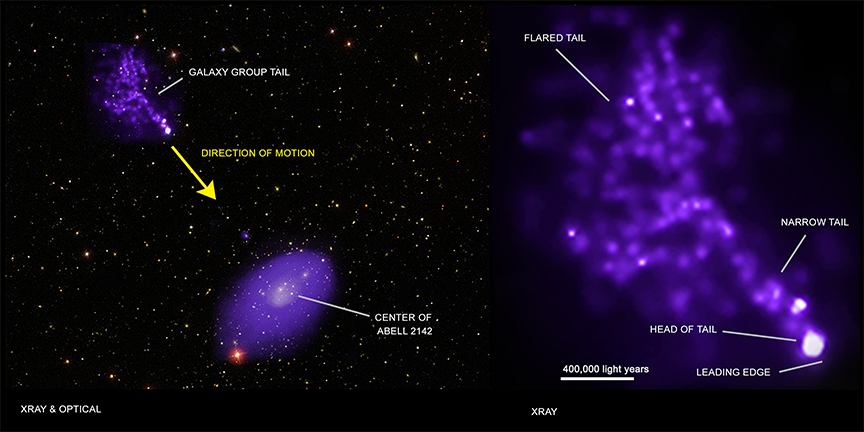}
\end{center}
\caption{The ram-pressure stripped tail of the A2142 infalling group, generating a cascade of turbulent motions with Mach numbers of 0.1--0.25 estimated from the surface brightness fluctuation analysis. Credit:X-ray: NASA/CXC/Univ. of Geneva, D. Eckert. Optical: SDSS provided by CDS through Aladin.
}
\label{f:tail}       
\end{figure}

\subsubsection{Further considerations based on X-ray morphology}\label{sect_xmorph}

In addition to analysing the power spectra of small-scale X-ray surface brightness fluctuations, the morphology and thermodynamical properties of specific substructures such as cold fronts or ram-pressure stripped tails seen on larger scales can provide additional information about the line-of-sight motions and the presence of turbulence or its suppression, or modification, by ICM viscosity, thermal conductivity, or magnetic fields. 

It was suggested early on that the ratio of thermal pressures at the stagnation point of a cold front and in the free stream region could be used to determine the velocity of the cold cloud with respect to the surrounding ICM \citep{vikhlinin2001}. This typically lead to near-sonic Mach numbers inferred from the observations \citep{vikhlinin2001,mazzotta2003,ohara2004,machacek2005}, although a few cases with small relative velocities (consistent with zero) were also identified \citep{markevitch2001,Tanaka2006}. The challenge here is that the gas parameters at the stagnation point usually cannot be measured directly, because the stagnation region is physically small and its X-ray emission is strongly affected by projection. Away from the stagnation point, the strong tangential shear expected to be associated with cold fronts should promote the growth of hydrodynamic instabilities which can be a source of turbulence in the ICM. Initially, cold fronts appeared remarkably sharp, both in terms of the density and the temperature jumps, suggesting that the cold front could be stabilised by a layer of magnetic field parallel to the discontinuity \citep[e.g.][]{vikhlinin2001b}, or viscosity \citep[e.g.][]{roediger2013b}. 
Recent, deeper observations have revealed evidence for the onset of Kelvin-Helmholz (KH) instabilities in a number of systems \citep{Roediger2012a496,roediger2012a,werner2016,walker2017}. Notably, in the cases of the cold fronts in A3667 and the Perseus Cluster, \citet{ichinohe2017} and \citet{ichinohe2018} calculated the thermal pressure deficit associated with candidate KH rolls compared to the ambient ICM; assuming this deficit is solely due to turbulent pressure support, they report $v_{\rm turb}\sim$300--400 km/s in both cases. Although these estimates are currently uncertain to within a factor of a few, this illustrates the power of combining high spatial resolution cold front morphology with direct measurements of line broadening from future high spectral resolution X-ray telescopes to constrain the microphysics of the ICM. Based on the onset of KH instabilities in A3667, \citet{ichinohe2017} put an upper limit on the effective ICM viscosity at 5\% of the isotropic Spitzer value.

Comparing tailored simulations and deep observations of the stripped galaxy M89 in Virgo, \citet{Roediger2015visc} and \citet{Kraft2017} do not find indications of viscous or other suppression or mixing of cold stripped gas with the hotter ICM in the wake of this galaxy. Similarly, the wake of NGC 1404 seems to be well mixed too, indicating that the effective viscosity is at most 5\% of the Spitzer value \citep{Su2017}.  However, to apply results from generic, idealised simulations to specific objects, the dynamical conditions of the stripped subcluster in question (e.g., the initial gas contents or the merger stage) need to be taken into account. \citet{eckert17b} used the X-ray surface brightness fluctuation analysis discussed in the previous subsection to study the kinematics within the tail and wake of the ram-pressure stripped group observed falling into Abell 2142 (Fig.\,\ref{f:tail}). The retrieved turbulent Mach numbers generated by this infall are in the range ${\rm Mach_{\rm 3D}}\sim 0.1$\,-\,0.25, which coincidentally seems to be similar to the level of subsonic turbulence driven in the ICM cores by the AGN feedback cycle. The shallow slope of the fluctuation power spectrum in Abell 2142 indicates that transport properties (such as thermal conduction) are far below the Spitzer value. 

\subsubsection{The kinematic Sunyaev-Zel'dovich effect}

In addition to the widely used thermal Sunyaev-Zel'dovich (SZ) effect, wherein cosmic microwave background (CMB) photons passing through the ICM are inverse Compton scattered from the low-frequency region of the blackbody spectrum to substantially higher energies, gas bulk motions in the ICM can lead to an additional distortion of the CMB spectrum \citep{sunyaev1980}, expressed as 
\begin{equation*}
\frac{\Delta T}{T} = - {\mbox{$\sigma_{\mbox{\tiny T}}$}} \int n_e \vec{\beta} d\ell.
\label{eq:dT_ksz}
\end{equation*}
Here, $\beta \equiv v \vec{z}/c$ is the unitless (bulk) velocity, and $\mbox{$\sigma_{\mbox{\tiny T}}$}$ is the Thompson cross-section. This kinematic SZ (kSZ) signal is subdominant to the thermal SZ, unless the gas velocity reaches a few tenths of a percent of the speed of light \citep[1000 km s$^{-1}$, see e.g.][]{Birkinshaw99}. In addition, since this effect manifests itself only as an overall shift in the CMB temperature, it is indistinguishable in principle from temperature variations corresponding to primary CMB fluctuations, making its direct detection rather challenging. However, on small angular scales where the contribution from primary CMB anisotropies becomes small, the kinematic SZ (kSZ) effect can allow measurements of the gas bulk velocities and especially velocity gradients in the ICM. 

The first spatially resolved map of the kSZ signal in a galaxy cluster was presented recently by \citet{Adam17}, using data obtained with the NIKA camera at the IRAM 30 m telescope. They observed the merging cluster MACS J0717.5+3745 and detected a dipolar structure in the kSZ signal, with peaks that correspond to two subclusters moving away and toward us at velocities of several thousand km/s. This opens the way for future high-resolution SZ observations promising to provide much more sensitive constraints on gas motions in the ICM via the kSZ effect.

\subsubsection{Optical/sub-mm line-emission} \label{sect_optical}

\begin{figure}
\includegraphics[width=0.5\textwidth]{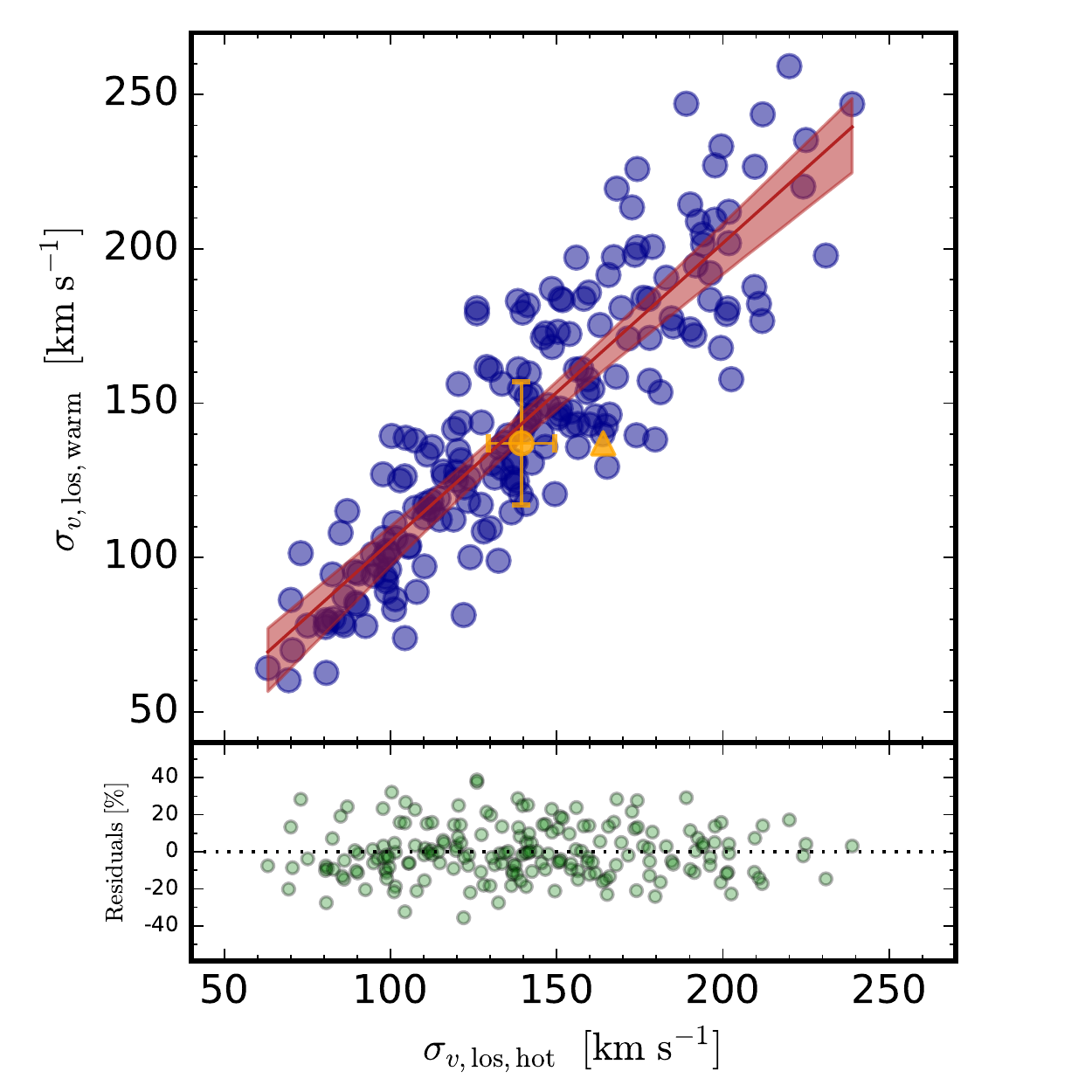}
\includegraphics[width=0.48\textwidth]{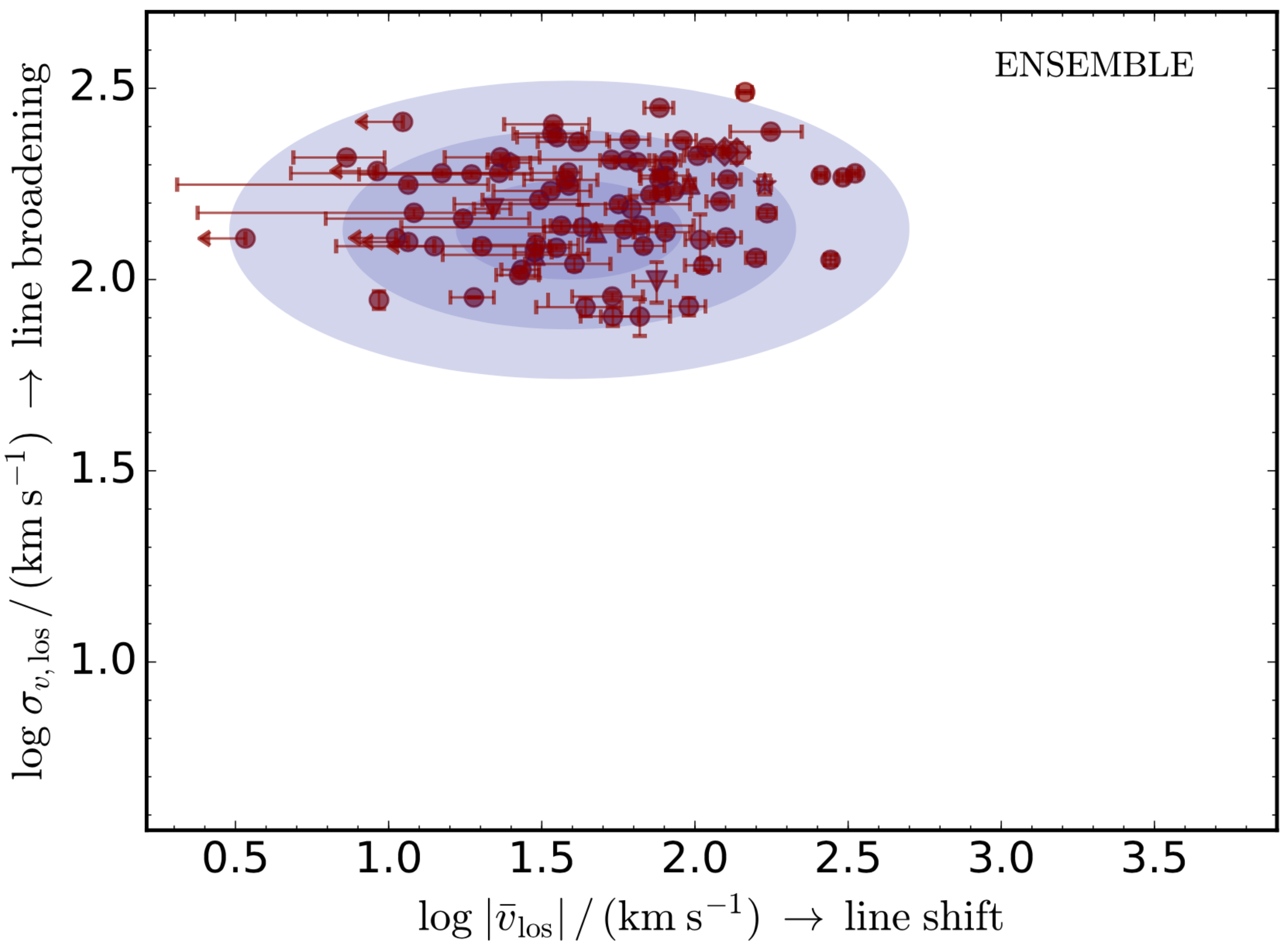}
 \vspace{0cm}
\caption{Left: Correlation
between the luminosity-weighted velocity dispersion of warm ($\sim 10^4$~K)  gas and the hot ICM (0.3\,-\,8 keV) in the cooling core ($\lesssim 50$ kpc). The red line and bands show the best-fit linear regression and the associated 99\% confidence interval. 
The yellow point shows the observational constraints from the Perseus cluster combining the SITELLE H$\alpha$+[NII] data with the {\it Hitomi} measurements (Section \ref{sect_hitomires}). 
Right: velocity dispersion versus the magnitude of the LOS velocity for the warm and cold gas phases, comparing the observational data (red points) with the predictions from numerical simulations (1-3$\sigma$ contours) for the ensemble detections. Adapted from \citet{gaspari2018}.}
\label{multiphase_corr}       
\end{figure}

Central cluster galaxies in cool core clusters are frequently observed to host optical line-emitting filaments \citep[see e.g.][and citations thereof]{crawford99}. More recently, these structures have been found to be multi-phase \citep[e.g.][]{edge2001, salome2006, sparks2009, salome2011, mittal2011, mittal2012, werner2014, tremblay2015, anderson2018}, consisting of gas from $<$100 K — 10$^{4}$ K as detected through molecular CO, ionised C and O, and Ly-$\alpha$ emission lines, much of which has been observed to be extended and filamentary. In particular, the high spatial and spectral resolution of ALMA has been instrumental in linking the warm ionized and cold molecular gas phases and uncovering the total mass (most of which is in the cold phase) within the filamentary systems \citep[e.g.][]{David14, McNamara14,Vantyghem16,Russell16,russell2017a,russell2017b, temi2017, Pulido18, simionescu2018,Tremblay18}. The excitation mechanisms of this multi-phase gas have been well studied, but are beyond the scope of this review.

The frequency of ionized filaments in the central galaxies of cool core clusters compared with those of non-cool core clusters imply they are connected with the hot cluster atmosphere either by cooling directly into filaments, being uplifted into the hot atmosphere from the central regions of the galaxy by the action of AGN feedback, or a combination of the two (see Section \ref{sec:AGNfeedback}). 
The morphologies and kinematics of these filaments can provide insight into the ICM velocity structure and physical properties \citep{hamer2016, mcdonald2012, Russell16, russell2017a,russell2017b, Tremblay18}. It was argued early on that the thin and long shape of ionised and molecular gas filaments implied a quasi-laminar or viscous flow with little small scale turbulence \citep[e.g.][]{Fabian2003a, Hatch2005}. However, initial debates regarding the origin of these structures made it unclear whether and to what extent their dynamics faithfully traced the motions of the ICM, or if these filaments were rather immune to their ambient medium like ``tree logs in the wind''. Recent ALMA observations have shown that lifting the observed amount of molecular gas out of the cluster cores requires an infeasibly efficient coupling between the molecular gas and the radio bubbles \citep{McNamara14,Russell16,russell2017b}; this, together with the fact that the velocities of the molecular gas clouds are well below their free-fall speeds, favour a scenario in which the observed filaments are the result of cooling instabilities in the X-ray gas triggered in situ by disturbances associated with the AGN feedback process \citep{gaspari13cca,voit2015,mcnamara2016}. If the line-emitting filamentary nebulae indeed originate from the cooling of the ICM, it is naturally expected that their velocity structure should trace that of the ambient X-ray plasma. 

By using high-resolution hydrodynamical simulations including self-regulated AGN jet feedback, \citet{gaspari2018} showed that all gas phases in cooling cores are tightly linked in terms of the {\it ensemble} LOS velocity dispersion (see Fig. \ref{multiphase_corr}), indicating that the spectroscopic measurements of the velocity dispersion in the warm/cold gas measured in wide (arcmin) apertures are reliable tracers of the velocity dispersion of the hot X-ray emitting gas. 
The retrieved best-fit correlation of the ensemble warm-gas velocity dispersion is  $\sigma_{v,{\rm warm}}=0.97^{+0.01}_{-0.02}\,\sigma_{v,{\rm hot}}+8.3^{+3.5}_{-5.1}$ (km~s$^{-1}$). The authors show that observations of massive galaxies and central group and cluster galaxies have little scatter in their ensemble line widths suggesting a common level of turbulence in the hot gas of 100--250 km/s, consistent with the direct line width measurement from hot gas in the Perseus cluster \citep{hitomi2016} and those of density fluctuations measured from X-ray images (see Section \ref{sect_fluct}). 
This study also showed that the opposite, pencil-beam approach is key to constrain the small-scale turbulence and bulk motions of single elements: the smaller the beam, the better we can probe the single cold/warm clouds raining via CCA toward the SMBH, which will trigger the subsequent feedback event. Most of the detected clouds and filaments have virial parameter $\gg 1$, i.e., they are highly dynamically supported as they inherit the velocity dispersion during the turbulence cascade from the parent hot halo \citep{gaspari2017}.

Another, potentially exciting, avenue to constrain gas motions of the ICM arises from heavy element hyperfine structure transitions in the radio probing 10$^{5–8}$ K \citep{sunyaev1984} and coronal emission lines \citep{anderson2016}. So called, due to their abundance in the optical spectra of the Sun, coronal lines originate from gas at $\sim$10$^{5–6}$ K. This phase of gas provides the missing link between the $\lesssim$10$^{4}$ K filaments and the hot ICM, offering a direct means of measuring the rate at which the ICM is condensing \citep{Cowie1981}. Moreover, the comparatively high spatial resolution of optical and radio measurements compared with the X-ray data would enable us to trace the hot gas velocities at sub-arcsecond scales. 

\citet{Graney1990}, \citet{Sarazin1991}, \citet{Voit1994} and most recently, using non-equilibrium cooling, \citet{Chatzikos2015} have modelled coronal-line emission in conditions appropriate to those of the cores of galaxy clusters. Despite evidence for gas at these temperatures from X-ray and UV lines \citep[e.g.][]{sanders2010b, pinto2014, Bregman2006, anderson2018}, hyperfine structure transitions of heavy elements have never been observed from cooling gas in galaxy clusters and optical coronal lines remain elusive with mostly upper limits reported \citep[e.g.][]{Hu1985, Heckman1989, Anton1991, Donahue1994, Yan1995}. \citet{Canning2011} report a 6.3 sigma detection of Fe X emission, a tracer of 10$^{5}$ K gas in the central galaxy of the Centaurus cluster with a velocity width of 300-650 km/s, though they conclude this is likely the result of gas heated by interaction with the radio lobes of the AGN rather than cooled from the hot ICM.

\subsection{Recent progress using micro-calorimeter X-ray spectroscopy}\label{sect_hitomires}

\begin{figure}[tbp]
\begin{center}
\includegraphics[clip=true,trim=0.0cm 0.0cm 0.0cm 0.0cm,height=2.2in]{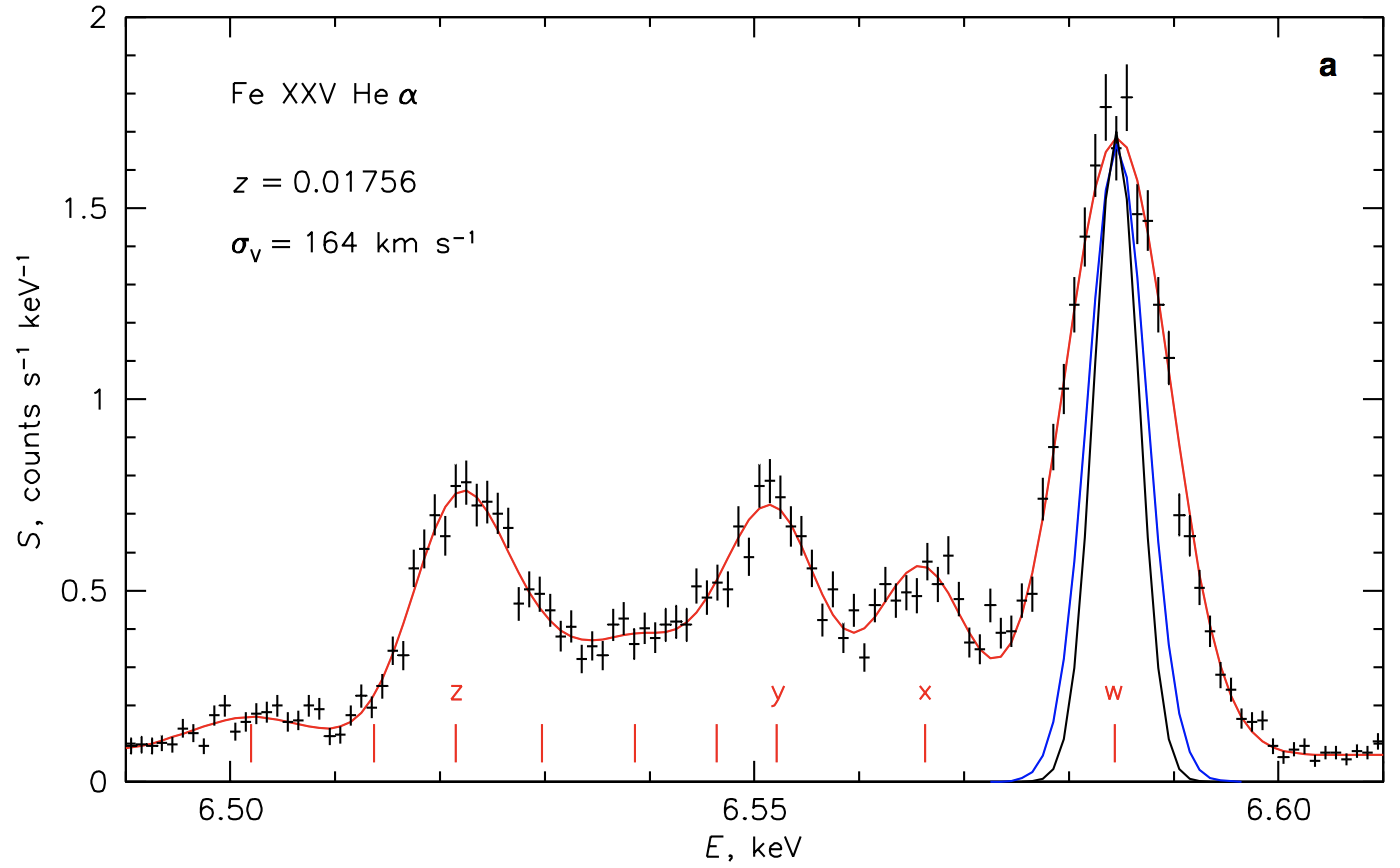}
\end{center}
\caption{Spectrum of the core of the Perseus Cluster obtained by the \textit{Hitomi} SXS, focusing on the Fe XXV He-$\alpha$ line complex. Instrumental broadening	with (blue line) and without (black line) thermal broadening are indicated, demonstrating that the line width due to turbulent velocities is clearly resolved for the first time. From \citet{hitomi2016}.}
\label{fig:sxsfek}
\end{figure}

The \textit{Hitomi} Soft X-ray Spectrometer (SXS) observations of the Perseus cluster allowed us to perform the first precise, direct measurements of velocity dispersions and bulk motions in the ICM. The exquisite spectral resolution of this detector (a factor of more than 20 better than conventional CCDs), combined with its non-dispersive nature that circumvents the line broadening for extended sources affecting the \emph{XMM-Newton} RGS, provided a break-through in our understanding of the dynamics of the hot, diffuse ICM. The line broadening due to turbulent velocities is clearly resolved, above the instrumental and thermal broadening (Fig \ref{fig:sxsfek}).

\begin{figure}
\begin{center}
  \includegraphics[width=5.3cm]{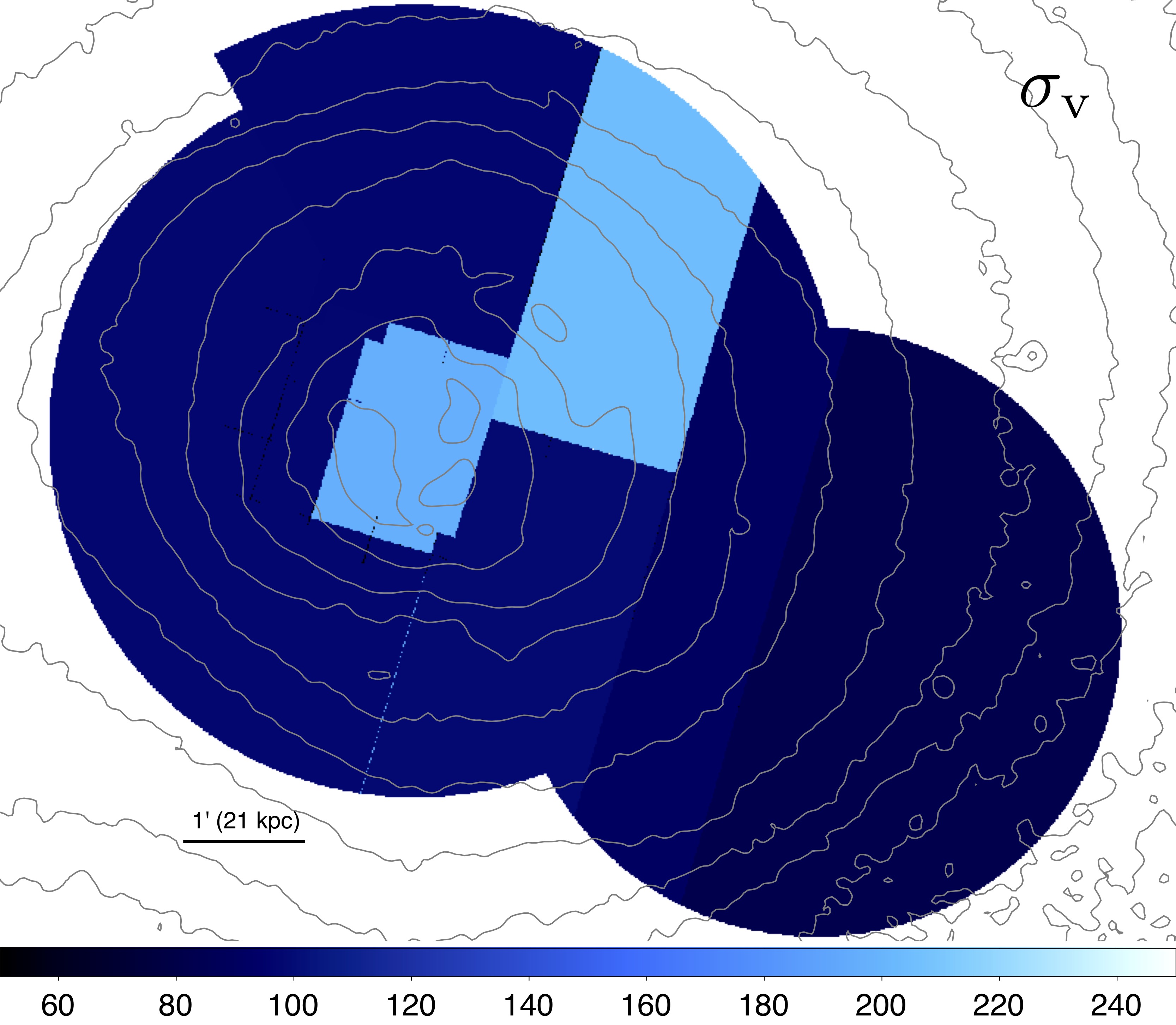}
  \includegraphics[width=5.3cm]{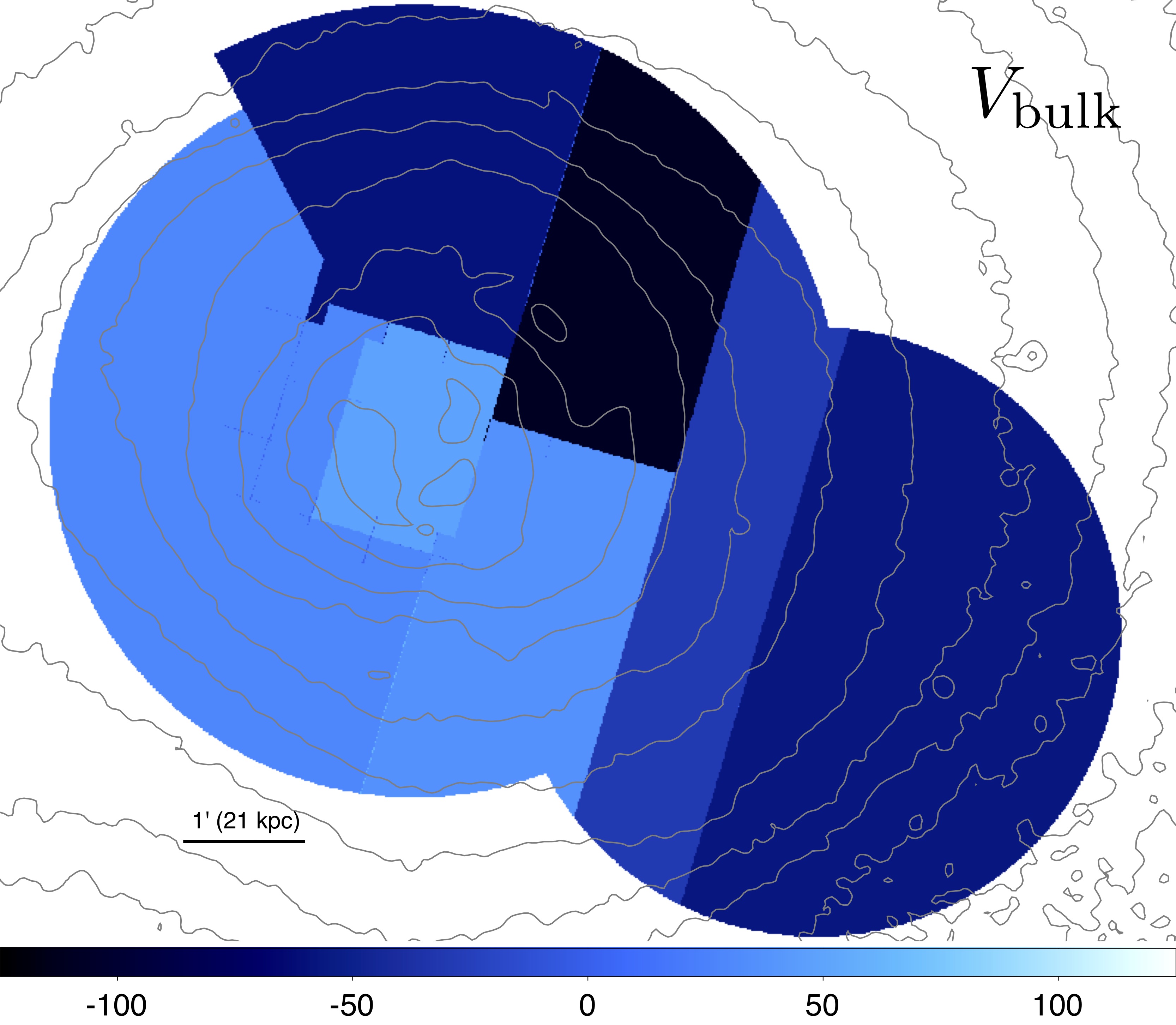}
 \end{center}
 \vspace{0cm}
\caption{PSF corrected maps of the velocity dispersion ($\sigma_{\rm v}$ , left) and bulk velocity ($v_{\rm bulk}$, right) in the core of the Perseus Cluster obtained from \textit{Hitomi}. The bulk velocity is expressed relative to $z = 0.017284$ (heliocentric correction applied). The unit of the values is km~s$^{-1}$. The \emph{Chandra} X-ray contours are overlaid. Adapted from \citet{HitomiV}.}
\label{hitomi_velocity}       
\end{figure}

The initial \textit{Hitomi} results yielded a velocity dispersion of $164\pm10$ km/s (at 90\% confidence) in a region located 30--60 kiloparsecs from the central nucleus, and a hint of a slightly higher turbulent velocity in the immediate vicinity of the AGN \citep{hitomi2016}. 
A subsequent, more in depth analysis presented in \citet{HitomiV} accounted for the smearing effects of the \textit{Hitomi} PSF and revealed that the velocity dispersion reaches maxima of approximately 200~km~s$^{-1}$ toward the central AGN and toward the AGN inflated north-western `ghost’ bubble (left panel of Fig.~\ref{hitomi_velocity}). Elsewhere within the observed region, the velocity dispersion appears constant around 100~km s$^{-1}$. There is a line-of-sight velocity gradient with a 100~km~s$^{-1}$ amplitude across the cluster core, consistent with large-scale sloshing of the core gas (right panel of Fig.~\ref{hitomi_velocity}). 

In principle, the observed line widths quoted above can be due both to small-scale near-isotropic turbulence and to a superposition of laminar motions along the line of sight (see discussion in Section 1.2). It is important to note therefore that, in addition to the measurements of line widths and shifts quoted above, the \textit{Hitomi} data allows for several other tests that can reveal the nature and driving scale of gas motions in the Perseus cluster core.

The first of these tests relies on measurements of the shapes of strong spectral lines \citep[as suggested by ][]{zhuravleva2011,shang2012,shang2013}. While turbulence on small scales will increase the observed line widths of well-resolved optically thin emission lines maintaining a Gaussian shape, gas motions on large scales will shift the line centroids. The superposition of large scale motions within the spectral extraction area should therefore lead to non-Gaussian features in the observed line profiles \citep[e.g.][]{inogamov2003}. \citet{HitomiV} report a lack of evidence for non-Gaussian line shapes, indicating that the observed velocity dispersion is dominated by small scale motions with a turbulent driving scale likely below 100 kpc, which is consistent with the size of the AGN jet inflated bubbles.

The second test relies on the suppression of lines with high oscillator strengths by resonance scattering \citep[][and Section \ref{res_scat}]{gilfanov1987,churazov2010b}. Line broadening and resonance scattering as a function of radius have a different dependence on the directionality of gas motions and therefore, in principle, they allow us to reveal and place constraints on the anisotropy of gas motions in bright cluster cores \citep{rebusco2008,zhuravleva2011}.

\begin{figure*}
\begin{minipage}{\textwidth}
\includegraphics[width=0.5\textwidth]{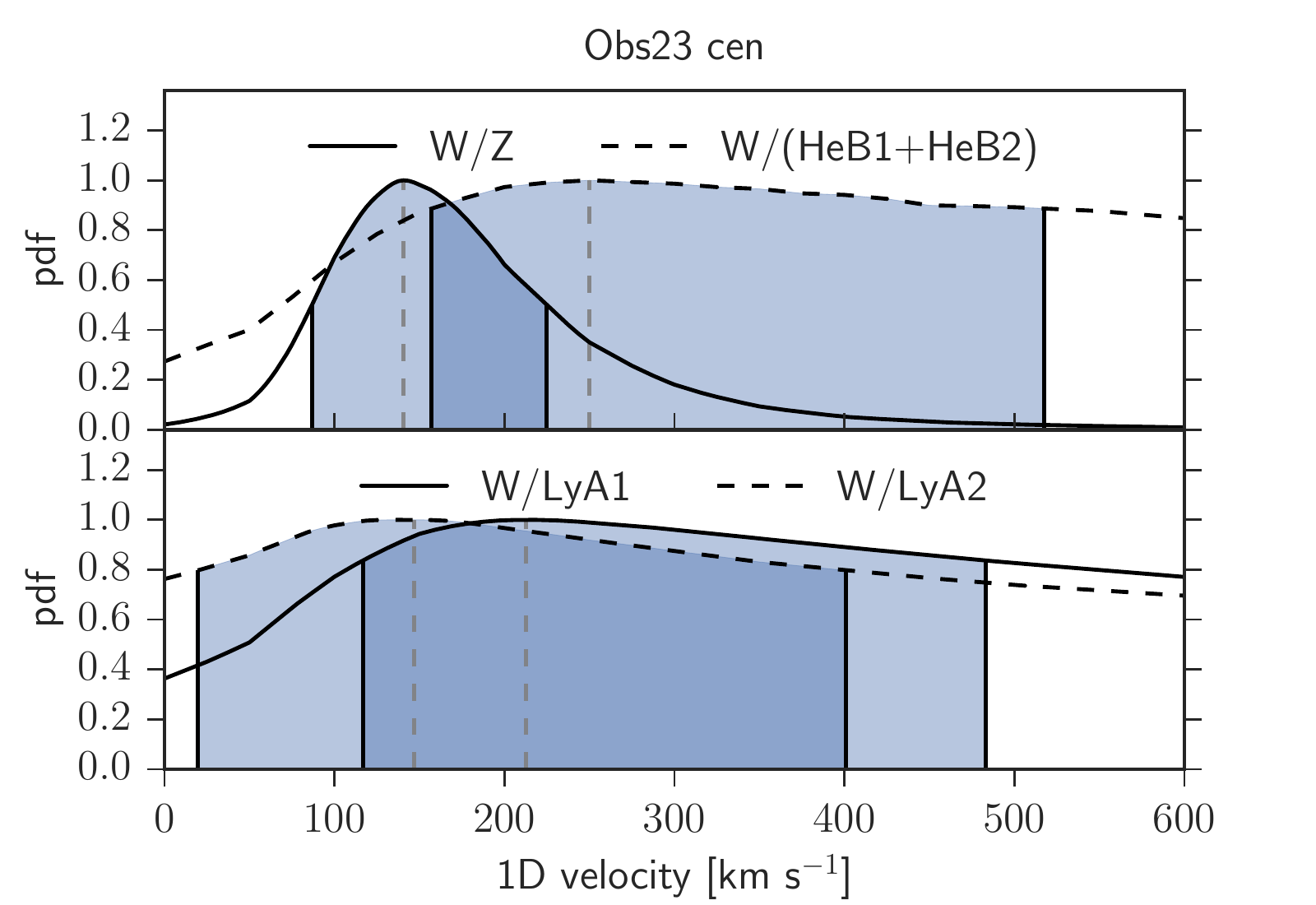}
\includegraphics[width=0.5\textwidth]{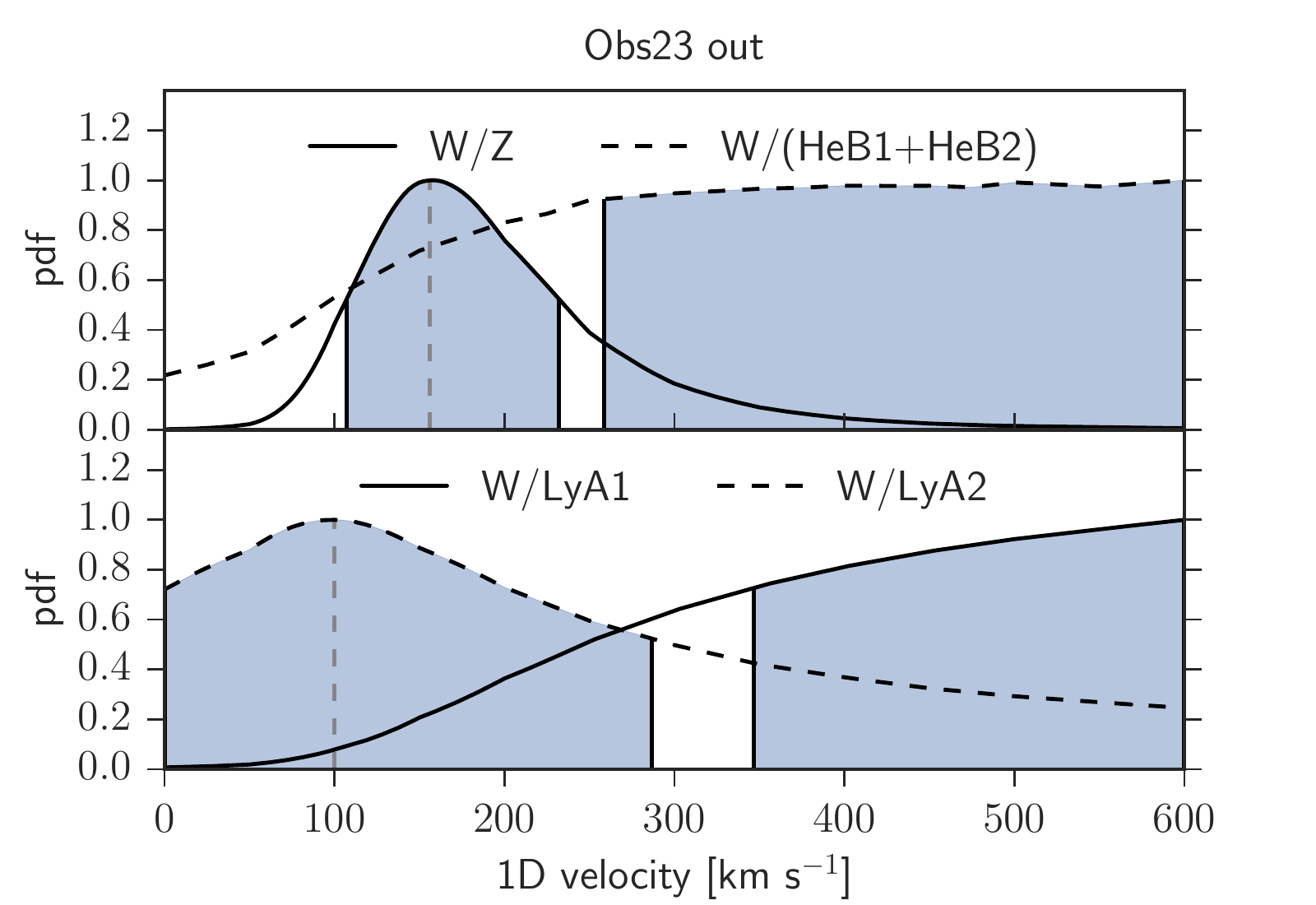}
\end{minipage}
\caption{Probability distributions for the turbulent velocities in the core of the Perseus Cluster inferred from measurements of the resonance scattering effect obtained by \textit{Hitomi}. The flux ratios of the resonant Fe XXV He$\alpha$ line (`W') with respect to various emission lines that are approximately optically thin (Fe XXV He$\alpha$ forbidden line `Z', Fe XXV He$\beta$, and two Fe XXVI Ly$\alpha$ lines) are compared to predictions from radiative transfer simulations to obtain these constraints. The left panel shows the inner $\sim$ 30 kpc region, while the right panel shows a region $\sim$ 30--100 kpc away from the cluster center. Especially in the outer region, uncertainties in the predictions of line emissivities in optically thin plasma models have a significant effect on the inferred turbulent velocities.
Adapted from \citet{HitomiRS}.
\label{fig:rsvel}
}
\end{figure*}

The {\it Hitomi} SXS observation of the Perseus cluster allowed the robust measurement of resonant scattering in the strongest X-ray line of the spectrum, the Fe XXV He$\alpha$ line (w). The measured flux suppression in this line is a factor of $\sim$ 1.3 in the inner $\sim$ 40 kpc, and decreases to $\sim$ 1.15 in the region $\sim$ 30-100 kpc away from the cluster center \citep{HitomiRS}. Comparing these suppressions with the results of radiative transfer simulations in the Perseus cluster, turbulent velocities of $150^{+80}_{-56}$ km/s and $162^{+78}_{-50}$ km/s have been measured in both regions (Fig. \ref{fig:rsvel}). The results are consistent with the direct measurements from line broadening \citep{HitomiV}, suggesting a lack of anisotropy in the gas motions (in agreement with the Gaussian shape of strong emission lines). However, both the systematic and the statistical uncertainties remain large. For instance, different choices of reference optically thin lines with respect to which the suppression of the Fe XXV He$\alpha$-w is calculated can affect the inferred turbulent velocities (Fig. \ref{fig:rsvel}). In order to provide meaningful constraints on the anisotropy of gas motions, both further improvements in the predictions of line emissivities in plasma models and deeper observations will be necessary. 

It is moreover interesting to compare the velocities obtained from the surface brightness fluctuation analysis and high-resolution \textit{Hitomi} spectra in Perseus, as a gauge for the validity of the method discussed in Section \ref{sect_fluct}. The total line-of-sight velocity dispersion observed by \textit{Hitomi} from the region $r \lesssim$ 100 kpc, can be estimated as $V_{\rm 1D}=\sqrt{\sigma^2+\Delta V ^2}$.  Here $\sigma$ and  $\Delta V$ are,  respectively, the line broadening measured in individual pixels and the centroid shift across the region \citep[see][for detailed discussion]{zuh2018}. Taking into account the radial distribution of the gas contributing to the line flux, one can interpret the \textit{Hitomi} value as an estimate of the velocity amplitude on scales of $\sim 60$~kpc. At this scale, the velocity dispersion measured by \textit{Hitomi} is in agreement with the fluctuations analysis amplitude \citep{Zhu14a, Zhu17,HitomiV}. Moreover, Figure \ref{multiphase_corr} reveals an excellent agreement between the \textit{Hitomi} velocity dispersion for the Perseus Cluster and the line width of the \textit{ensemble} warm H$\alpha$+[NII] phase obtained from SITELLE data \citep{gaspari2018}. In a subsequent work, \citet{gendron2018} studied several smaller-scale patches of the NGC 1275 nebula detected with SITELLE, finding that, while the velocity dispersions of the filamentary nebula surrounding NGC 1275 are indeed as low and uniform as the \textit{Hitomi} measurements of the ICM, there is no correlation between the \textit{bulk} velocities of the different phases.

\section{Implications for distinguishing the source of turbulent motions and constraining the ICM viscosity}\label{sect_implic}

The publication of the precise measurements of the ICM dynamics in the core of the Perseus Cluster obtained by \textit{Hitomi} has sparked a lively debate in the community, with questions revolving around 
\begin{enumerate}
\item what is the main physical mechanism that is driving the gas motions?
\item is the observed level of turbulence sufficient to balance radiative cooling?
\item what, if any, are the physical implications for constraining the effective viscosity of the ICM?  
\end{enumerate}
In this section, we summarise and discuss the various proposed interpretations.

In terms of the source of gas motions, it is clear from Section \ref{sect_theor} that we are looking at a superposition of various physical processes which all predict roughly similar turbulent velocities within the central region of clusters (at the level of a few hundred km/s). Based on the average level of turbulent motions alone, it is then difficult, if not impossible, to distinguish which of these different processes is the dominant one. This question can be addressed through a spatially-resolved comparison of the line shifts and widths as a function of radius or in regions of various X-ray morphology (i.e. associated with cavities or aligned with existing cold fronts).

The peaks in $\sigma_v$ seen in the \textit{Hitomi} observations of the Perseus Cluster appear to indicate that gas motions are driven both at the cluster center by the current AGN inflated bubbles and by the buoyantly rising ghost bubble with a diameter of $\sim 25$~kpc \citep{HitomiV}. This appears to contradict models in which gas motions are sourced only at the center (during the initial stages of bubble inflation) or only by structure formation. 

At the same time, surface brightness features in the Perseus cluster indicate the presence of gas sloshing \citep{churazov2003,simionescu2012,walker2017} which will also drive a part of the observed motions (see Section \ref{sec:sect_cf}). The relative uniformity of the velocity dispersion, apart from the regions associated with AGN bubbles, may be consistent with sloshing-induced turbulence \citep{zuhone2013}. In this case, the shearing motions associated with gas sloshing are also expected to contribute to the velocity dispersion observed throughout the investigated area and thus the true turbulence might be even weaker. The spatial coverage of the \textit{Hitomi} pointings, as well as the spatial resolution, limits our ability to fully disentangle the impact of the AGN and ongoing gas sloshing on the velocity structure of the ICM. Using the density fluctuations in deep \emph{Chandra} observations to estimate the turbulent velocities instead, \citet{walker2018} suggest that these are broadly consistent with being produced by sloshing alone outside the central 60 kpc. 

The problem becomes even more complex when the expected velocity field from cosmological large-scale structure formation is taken into account, in addition to AGN feedback and idealised gas sloshing.
\citet{lau2017} performed an analysis of mock \textit{Hitomi} observations of clusters from cosmological simulations and isolated clusters with AGN feedback (from \citealt{gaspari2012a}). They concluded that cosmic accretion and mergers could produce line-of-sight velocity dispersions and line shifts compatible with the \textit{Hitomi} observations of Perseus, while AGN feedback is able to produce velocity dispersion measurements which are compatible with the \textit{Hitomi} observations, but not the core-scale velocity gradient, since the turbulence driven by AGN feedback is too stochastic. \citet{bou2017} reached similar conclusions using mock \textit{Hitomi} observations of simulated isolated clusters with AGN-driven jets and models which included realistic cosmic substructure: AGN feedback can produce the observed line widths, but producing the observed line shifts requires gas motions driven by mergers. 

Spatially-resolved high-resolution spectroscopy covering a larger area of the Perseus Cluster core, as well as a number of other systems, will be crucial to reach an agreement on which physical process is the main source of gas motions, and how this changes as a function of radius.

In terms of the feasibility of balancing radiative cooling through turbulent dissipation, \citet{Zhu14a} first showed, using the surface brightness fluctuation analysis, that turbulent heating is not only sufficient to offset radiative cooling but indeed appears to match it locally at each radius, for 7 different annuli in Perseus and 4 in M87. The fact that the turbulent velocities inferred from this surface brightness fluctuation analysis were found to be in very good agreement with the direct constraints from the \textit{Hitomi} SXS (Section \ref{sect_hitomires}) lends further support to this interpretation. On the other hand, \citet{fabian2017} and \citet{Bambic18} argue that this low-velocity turbulence cannot spread far across the cooling core during the fraction of the cooling time in which it must be replenished, and hence conclude that another heating mechanism such as sound waves must dominate. However, this argument can be circumvented if the turbulence is driven in situ by the rising bubbles, as suggested by the higher velocity dispersion associated with the ghost cavity in the Perseus Cluster \citep{HitomiV}. Ultimately, tests based on the radial dependence of the line widths and line shapes discussed in Section \ref{sec:AGNfeedback} are expected to shed more light on this issue in the future.

\begin{figure}
\begin{center}
\includegraphics[width=0.48\textwidth]{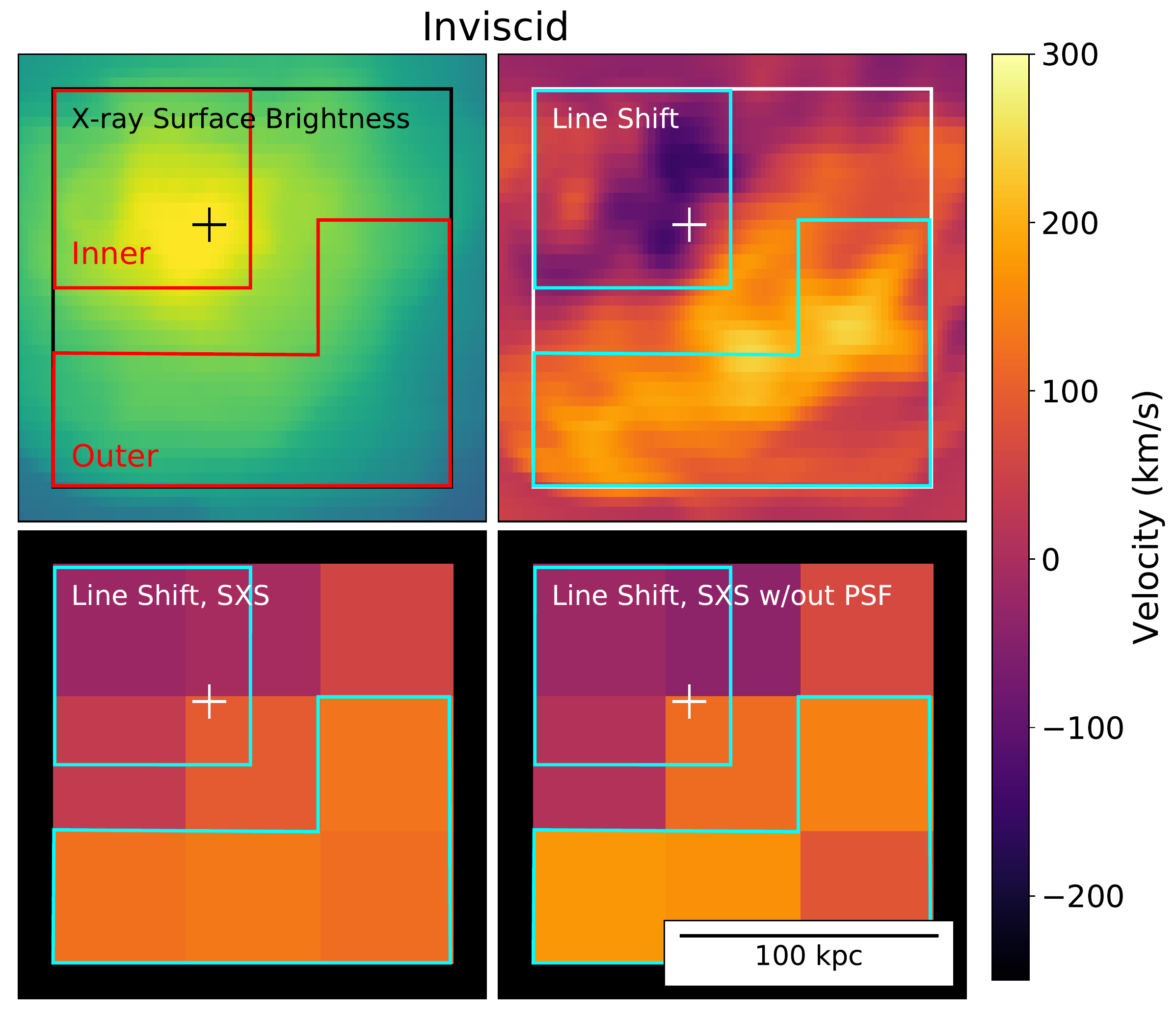}
\includegraphics[width=0.48\textwidth]{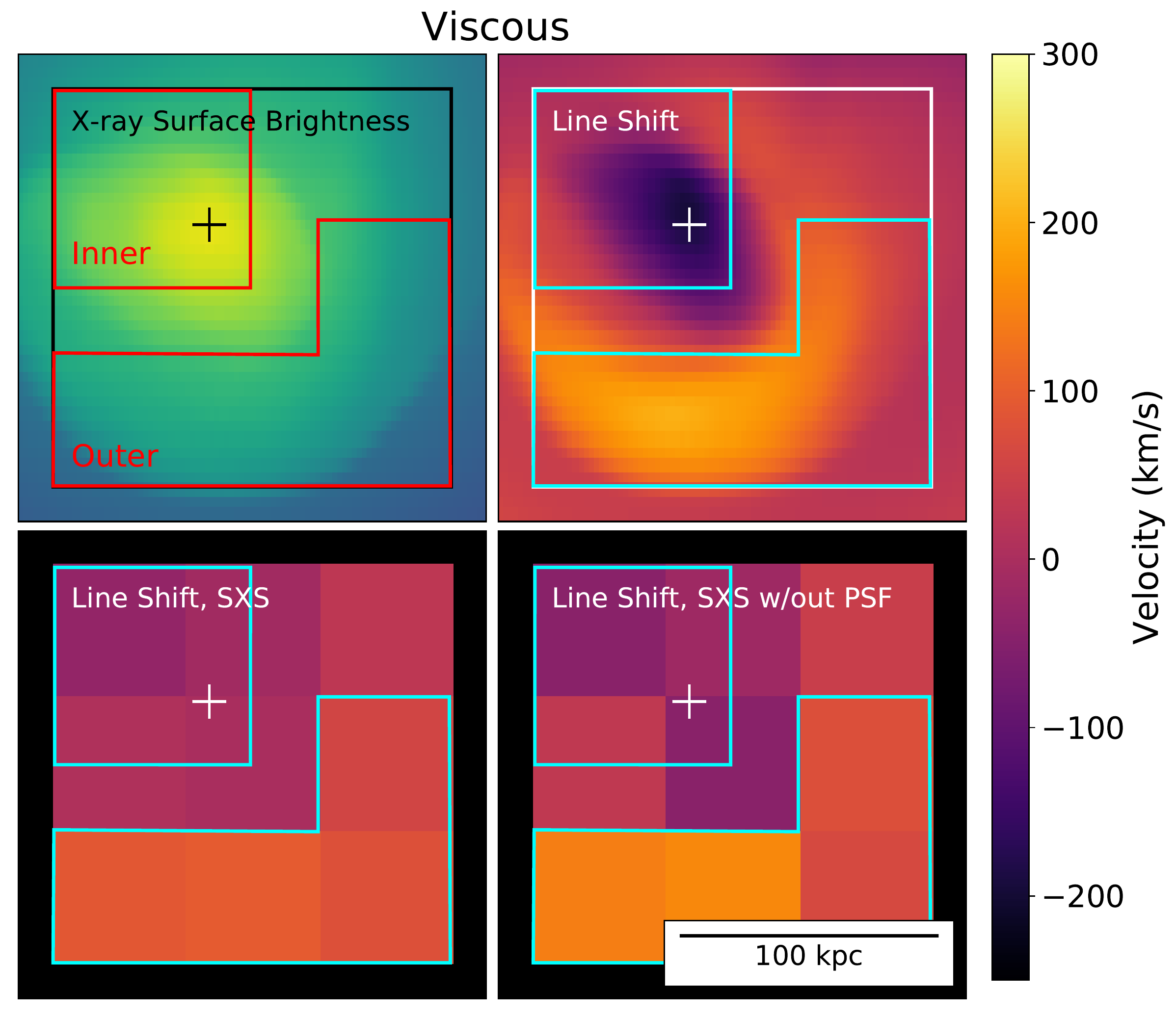}
\caption{Mock line shift maps for a Perseus-like cluster from \citet{zuh2018} from two hydrodynamic simulations, one inviscid and the other highly viscous. The top panels show the full simulation spatial resolution while the bottom show the {\it Hitomi} angular resolution. It is not possible to discern the difference in the velocity fields due to viscosity in the mock {\it Hitomi} observations.\label{fig:mock_hitomi}}
\end{center}
\end{figure}

In terms of the microphysical properties of the ICM, it is at first sight tempting to interpret the relatively small gas velocities measured by \textit{Hitomi} in a cool core cluster hosting strong signatures of sloshing activity as evidence for a high effective viscosity of the X-ray plasma. Indirect probes of the gas velocities in other cool core clusters, such as surface brightness fluctuations or the width of spectral lines from the warm (rather than hot) phase, similarly support relatively low turbulent velocities, as discussed above. To investigate this, \citet{zuh2018} performed two simulations of gas sloshing in a galaxy cluster core similar to the Perseus cluster, in the absence of AGN feedback, to study the effects of sloshing in isolation on the line-of-sight measurements of the velocity field; the simulations differed only in that the first was inviscid and the second was highly viscous with the isotropic Spitzer viscosity. \citet{zuh2018} found an orientation of the line of sight angle which reproduced both the spiral shape of the cold fronts and the velocity shear observed by \textit{Hitomi} across the core region, and performed mock \textit{Hitomi} observations. These observations were used to produce measurements of the line shift and width similar to those obtained for Perseus. Their results showed that at the $\sim$1' spatial resolution afforded by \textit{Hitomi}, it is very difficult to distinguish the velocity field of a highly viscous ICM from an inviscid one (see Figure \ref{fig:mock_hitomi}). This is because the broadening of spectral lines is produced mainly by nearly core-size eddies which are only marginally affected by even the high level of viscosity employed in the simulation \citep{zuh2016}. 

In addition, as mentioned above, numerical models of cosmic accretion and AGN feedback presented by \citet{lau2017} and \citet{bou2017} are able to roughly reproduce the level of turbulent motions observed by \textit{Hitomi} without including viscosity explicitly in the calculation. Together with the \citet{zuh2018} results, this not only implies that gas motions from cosmic accretion/mergers/sloshing on the one hand and AGN feedback on the other are complementary drivers of the velocity field in cluster cores, but viscosity is not required to explain the level of gas motions seen in Perseus, provided that the system has not experienced a major merger recently. This is in line with the results discussed in Section \ref{sect_xmorph}, where the morphology of cold fronts and ram-pressure stripped tails suggest a low viscosity of the ICM for the cases of M89 \citep{Roediger2015visc,Kraft2017}, NGC1404 \citep{Su2017}, A2142 \citep{eckert17b}, and A3667 \citep{ichinohe2017}. 

Constraints on the microphysics of the ICM will require both a larger sample of clusters for which precise velocity measurements are available, but also an ensemble of observational strategies; while X-ray line broadening alone may be insufficient to obtain an estimate of the plasma viscosity, combining this information with measurements of resonance scattering and high-spatial resolution analysis of the X-ray morphology will provide a very powerful technique accessible to future missions, as discussed below.

\section{Future prospects}\label{sect_future}

Future X-ray missions, like {\it XRISM}, {\it Athena} and {\it Lynx} will measure radial profiles of the gas velocity dispersion and bulk velocities in many objects, allowing us to probe both the velocity amplitudes and scales \citep{Zhu12,zuh2016,roncarelli18}, and, therefore, systematically compare the imaging and spectral techniques. We summarise below the progress expected from these future missions, together with the improvements expected for experiments in other wavebands.

\subsection{Measuring gas motions with \emph{XRISM}}

The X-Ray Imaging and Spectroscopy Mission (\emph{XRISM}), formerly \emph{XARM} (the X-ray Astronomy Recovery Mission, \citealt{XARM}) is planned as a successor of the \textit{Hitomi} satellite, and will carry a high spectral resolution X-ray microcalorimeter (\emph{Resolve}), which is identical to the SXS. This will allow to extend detailed studies like that performed by \textit{Hitomi} for the Perseus Cluster to many other systems, and to softer X-rays (the SXS was planned to cover energies down to 0.3~keV, but was limited to $>$ 2~keV because of the transmission of the Beryllium filter that was in place during the early phase of the operation). The progress expected for galaxy cluster observations, including the measurement of gas motions in the ICM, using \emph{Resolve} is very similar to the science case for the SXS, summarised in \citet{kitayama2014}.

In terms of constraining the nature and origin of gas motions, as well as the microphysical properties of the ICM, \emph{XRISM} is relevant for a number of complementary aspects. Firstly, \emph{Resolve} will allow us to study the gas velocities in a larger sample of relaxed, cool-core clusters, and evaluate the impact of the central SMBH on the ICM dynamics. In a few cases (most notably, M87), the X-ray surface brightness features associated with the ongoing AGN feedback can be spatially resolved, and differential tests of the nature of gas motions corresponding to and offset from these AGN-related substructures can inform us about the mechanisms through which supermassive black holes drive turbulence in the ICM. 

Secondly, \emph{Resolve} will provide further insights into the impact of minor and major mergers in determining the kinematics of the ICM. Differential tests of the velocity structure performed for regions located in- and outside cold fronts, or on/offset from bright ram-pressure stripped tails of infalling subhaloes will probe the importance of minor mergers. Major mergers of galaxy clusters will also be a promising target, due to the fact that they produce bulk motions up to several 1000~km~s$^{-1}$ and turbulent motions of $\sim$1000~km~s$^{-1}$. The strongest and the most interesting signals will be observed for mergers occurring predominantly along the line of sight. Due to the complex state of gas motions in mergers and projection effects (see Section \ref{sec:sect_cf}), all these observations will have to be interpreted very carefully in combination with simulations. 

\emph{XRISM} measurements will likely focus on the inner regions of nearby massive galaxy clusters. A detailed characterisation of the gas velocity profile out to beyond $r\approx R_{2500}$ will require a significant investment of observing time, and can only be performed for a very limited sample of objects. However, if this time investment is made, \citet{nagai13} show that the analysis of mock SXS (analogous to \textit{Resolve}) spectra of a relaxed galaxy cluster extracted from cosmological numerical simulations recovers the 3D (deprojected) mass-weighted velocity dispersion profile up to $r\approx R_{500}$ accurately, while slightly ($\sim 30-50$~~km~s$^{-1}$) underestimating the \textit{projected} mass-weighted gas velocity dispersion in a given radial bin. This difference occurs because the measured velocity is spectral-weighted, and hence the inner regions where the gas density is higher but the gas velocity is smaller carry a higher weight. Going forward, this synergy between numerical simulations, mock observations, and real data will need to be employed frequently in order to correctly interpret the measurements and uncover any potential biases or significant projection effects \citep[see also][]{ota2018}.

To effectively expand our knowledge of the ICM velocity structure towards the cluster outskirts, future missions beyond \emph{XRISM} will be necessary. In addition to the significant progress expected to be obtained with \emph{Athena} (see the next subsection), the Chinese-led Hot Universe Baryon Surveyor (HUBS\footnote{\href{http://heat.tsinghua.edu.cn/~hubs/en/index.html}{http://heat.tsinghua.edu.cn/\~{}hubs/en/index.html}}) is another promising mission concept that is currently under consideration. With a large effective area ($\sim$1000 cm$^2$) and large field of view ($\sim$ 1 square degree), moderate angular resolution of 1 arcmin, and high spectral resolution (2 eV at 0.6 keV) in the soft X-ray band (0.1 to 2 keV), HUBS is optimised to look for the ‘missing baryons’ in intergalactic and circumgalactic space. However, its capabilities are also very well suited to provide robust measurements the widths and shifts of X-ray lines  in the faint outskirts of nearby clusters of galaxies (see also the related chapter on ``The physics of galaxy cluster outskirts'', \citealt{WalkerISSI}, this volume). 

\subsection{Measuring gas motions with \textit{Athena}/X-IFU}

\emph{Athena} \citep{nandra13} is the second large mission selected in the framework of the European Space Agency (ESA) \emph{Cosmic Vision} program and it is scheduled for launch in the early 2030s. The mission will carry a high-throughput X-ray telescope with an unprecedented effective area of 1.4 $m^2$ at 1 keV and an angular resolution of 5 arcsec. The telescope will feed two instruments in the focal plane, the Wide-Field Imager (WFI) and the X-ray Integral Field Unit (X-IFU). The X-IFU \citep{barret18} is an array of $\sim4,000$ transition-edge sensors (TES) with a spectral resolution of 2.5 eV at 7 keV and covering a field of view of 5 arcmin diameter. The capabilities of X-IFU will allow us to detect line shifts and broadening down to $\sim20$ km/s and map them at a resolution of 5 arcsec (see also \citealt{ettori13}).

\begin{figure*}[!ht]
\includegraphics[width=\textwidth]{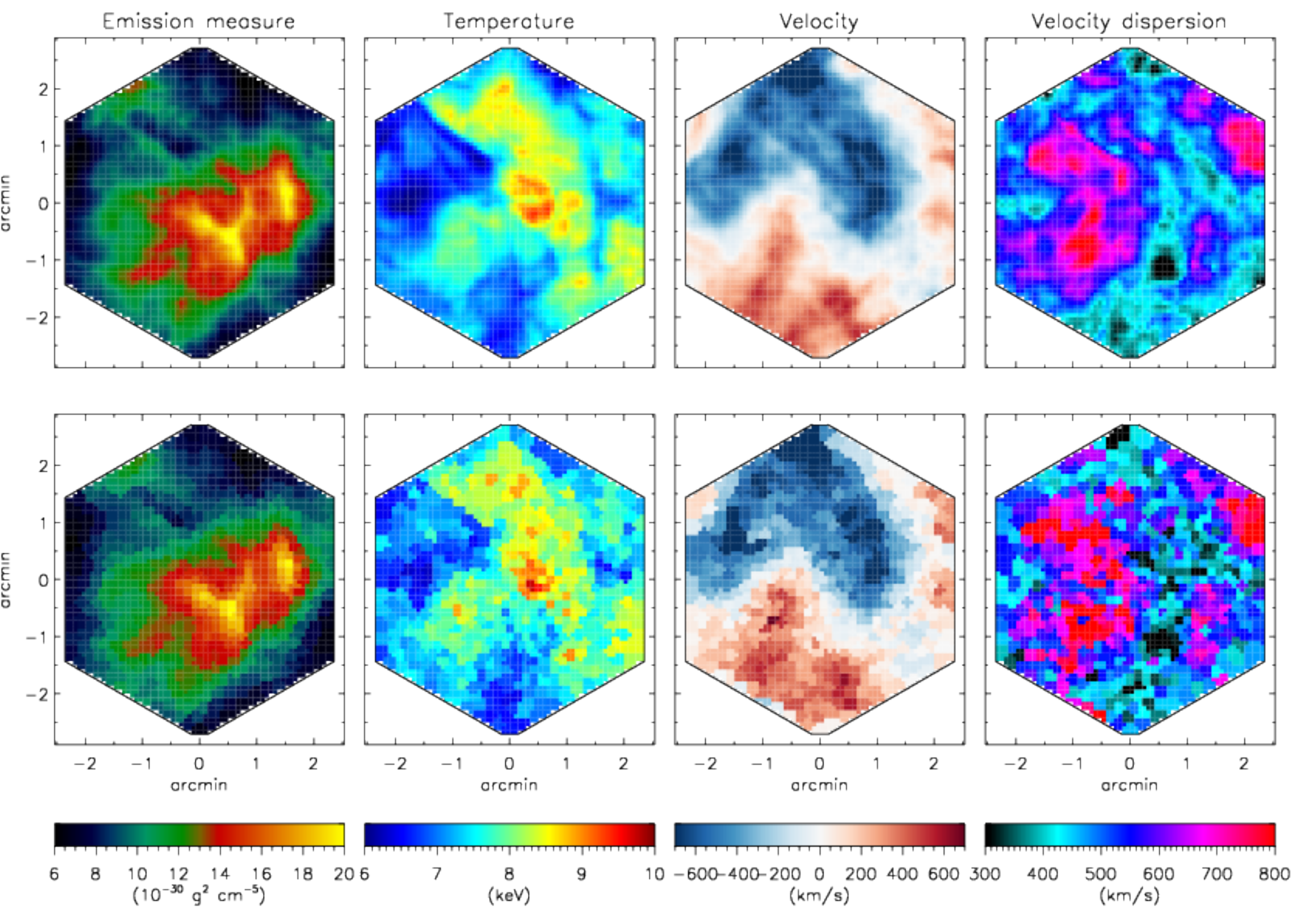}
\caption{True (top) and reconstructed (bottom) images of gas emissivity (far left), temperature (center-left), bulk velocity (center-right), and velocity dispersion (far right) for a fiducial Coma-like simulated cluster (\citealt{Gas13}) observed with the \textit{Athena}/X-IFU. Adapted from \citet{roncarelli18}.
\label{fig:XIFU}
}
\end{figure*}

To test the capabilities of \emph{Athena}/X-IFU at recovering gas motions in galaxy clusters, \citet{roncarelli18} presented mock X-IFU observations of a Coma-like cluster simulated at high resolution with the hydrodynamical code FLASH4 \citep[see][for details]{Gas13}. Three-dimensional data cubes extracted from the simulation were then fed into the SIXTE observation simulator \citep{wilms14} to create a long (1 Ms) mock observation of the fiducial cluster placed at $z=0.1$, including instrumental effects such as instrument response, point-spread function, and particle background. The mock observation was then analyzed by fitting $\sim600$ spectra binned to reach a target of 30,000 counts per region, and maps of the main quantities of interest (emission measure, temperature, bulk velocity, and velocity dispersion) were reconstructed. 

In Fig. \ref{fig:XIFU} (from \citealt{roncarelli18}), the reconstructed maps (bottom panels) compared to the true, input maps (top panels) are shown. The remarkable ability of X-IFU to reconstruct accurately the properties of the simulated cluster is evident. \citet{roncarelli18} further constrained the statistical accuracy and bias, if any, in the reconstructed measurements. In particular, the true bulk velocity of the simulated cluster can be reconstructed with a very high level of accuracy. The reconstructed velocity correlates with the true one with a Spearman correlation coefficient $r_s=0.960$. A clear correlation between reconstructed and true values is also found for the velocity dispersion ($r_s=0.765$), although the statistical uncertainties are larger. In both cases, biases in the reconstructed values were found to be at the level of 5\% or less. Overall, the simulations presented in \citet{roncarelli18} demonstrate that \emph{Athena}/X-IFU will allow us to perform accurate and spatially-resolved observations of gas motions in the ICM, thus enabling a leap forward in our understanding of these phenomena.

The most direct estimate of the ICM viscosity would require a measurement of the dissipation scale of turbulence from the power spectrum of measured line shifts. This scale is of the order of $\sim$1-10~kpc, for reasonable estimates of the ICM turbulent properties and the assumption that viscosity arises from ion collisions. \emph{Athena}'s 5 arcsec resolution may reach for the first time the dissipation scale of turbulence, thus placing direct constraints on the ICM viscosity. However, if this viscosity is very low, a definitive measurement may require the $\sim$1'' angular resolution of {\it Lynx}, a NASA large class X-ray mission concept being considered for the 2020 decadal survey \citep{Gaskin2017}.

\subsection{Synergies beyond the X-ray band}

Complementary to X-ray studies, subarcminute resolution multi-frequency SZ experiments such as (but not limited to) upgrades to ALMA, MUSTANG-2, NIKA2, and CONCERTO will greatly enhance our capability to measure the product of velocity and electron opacity of the cluster gas through the kinetic SZ effect. For further details, we refer the reader to the dedicated review on `Astrophysics with the Spatially \& Spectrally Resolved Sunyaev-Zel’dovich Effect' in this volume \citep{SZ_ISSI}. Comparisons between observations of bulk velocities of the gas from X-ray and high-resolution kinematic SZ measurements and galaxy velocities in the optical and infrared could potentially probe the collisionality of dark matter, assuming the galaxies serve as a reliable tracer of the kinematics of the latter. Leveraging the high spectral resolution of current/next-generation radio and optical/IR telescopes (e.g., ALMA, MUSE, JWST, SKA, ELT) will be key to unveil the kinematics of the ensemble warm filaments and cold clouds, and perform detailed comparisons between the velocity structure of this colder phase and that of the ICM/IGrM in cool-core systems. 

Synergy with observations in other wavebands is also very promising for major mergers between clusters of galaxies. Strong gas turbulence is believed to be the driver for radio halos, produced by the reacceleration of cosmic-ray electrons in the cluster magnetic field (see related chapter by \citealt{vanWeerenISSI} in this volume). X-ray measurements of line broadening in clusters, in combination with powerful new radio data obtained from LOFAR and the JVLA, will determine if the level of turbulence in the ICM is strong enough to produce the observed radio emission.

\begin{acknowledgements}
A.S. gratefully acknowledges support by the Women In Science Excel (WISE) programme of the Netherlands Organisation for Scientific Research (NWO). 
M.G. is supported by NASA through Einstein Postdoctoral Fellowship Award Number PF5-160137 issued by the \emph{Chandra} X-ray Observatory Center, which is operated by the SAO for and on behalf of NASA under contract NAS8-03060. Support for this work was also provided by \emph{Chandra} grant GO7-18121X.
D.N.\ acknowledges Yale University for granting a triennial leave and the Max-Planck-Institut f\"ur Astrophysik for hospitality when this work was carried out. 
N.W. is supported by the Lend\"ulet LP2016-11 grant awarded by the Hungarian Academy of Sciences.
E.R. acknowledges the support of STFC, through the University of Hull’s Consolidated Grant ST/R000840/1 and access to viper, the University of Hull High Performance Computing Facility.
\end{acknowledgements}

\bibliographystyle{aps-nameyear}      
\bibliography{gasmotions}    

\end{document}